\documentclass[11pt]{article}
\usepackage{latexsym}
\usepackage{graphicx}
\usepackage{amssymb,amsmath}
\usepackage{amsfonts}   

\def\fig#1{Fig.~\ref{#1}}
\def\figs#1#2{Figs.~\ref{#1} and \ref{#2}}

\def\e3{$\epsilon_3$}

\def\ch2{$\chi^2$}

\def\co#1{{\ifmmode{\cal O}_{#1}\else${\cal O}_{#1}$\fi}}

\newdimen\unit
\def\point#1 #2 #3{\vbox to0pt{\kern-#2\unit
 \hbox{\kern#1\unit#3}\vss}
\nointerlineskip}
\newcommand\newc{\newcommand} 

\newcommand{\be}{\begin{equation}}
\newcommand{\ee}{\end{equation}}
\newcommand{\bea}{\begin{eqnarray}}
\newcommand{\eea}{\end{eqnarray}}

\newcommand{\gev}{\mbox{ GeV}}
\newcommand{\tev}{\mbox{ TeV}}

\newcommand{\cl}{\text{CL}}

\newcommand{\alphaemmz}{\alpha_{\text{em}}(M_Z)^{\overline{MS}}}
\newcommand{\alphas}{\alpha_s(M_Z)^{\overline{MS}}}
\newcommand{\like}{\mathcal{L}}

\newc\anap[3] 
                 {{\em Astron. Astrophys.} {\bf #1} (#2) #3} 
\newc\ap[3] 
                {{\em Ann. Phys.} {\bf #1} (#2) #3}
\newc\apj[3] 
                {{\em Astrophys. J.} {\bf #1} (#2) #3}
\newc\app[3] 
                {{\em Astropart. Phys.} {\bf #1} (#2) #3}
\newc\aps[3] 
                {{\em Astrophys. J. Suppl.} {\bf #1} (#2) #3}

\newc\arnps[3] 
                {{\em Ann. Rev. Nucl. Part. Sci.} {\bf C#1} (#2) #3}
\newc\cpc[3] 
                {{\em Comp. Phys. Comm.} {\bf #1} (#2) #3}
\newc\epj[3] 
                {{\em Eur. Phys. J.} {\bf C#1} (#2) #3}

\newc\jcap[3] 
                {{\em JCAP} {\bf #1} (#2) #3}
\newc\jhep[3] 
                {{\em JHEP} {\bf #1} (#2) #3}
\newc\ijmp[3] 
                {{\em Int. J. Mod. Phys.} {\bf A#1} (#2) #3}
\newc\mnras[3] 
                 {{\em Mon. Not. Roy. Astron. Soc.} {\bf #1} (#2) #3}
\newc\nca[3] 
                {{\em Nuovo Cimento} {\bf #1} (#2) #3}
\newc\nim[3] 
                {{\em Nucl. Instrum. Methods} {\bf #1} (#2) #3}
\newc\nima[3] 
                {{\em Nucl. Instrum. Methods} {\bf A#1} (#2) #3}
\newc\nat[3] 
                {{\em Nature} {\bf #1} (#2) #3}
\newc\npb[3] 
                {{\em Nucl. Phys.} {\bf B#1} (#2) #3}
\newc\plb[3] 
                {{\em Phys. Lett.} {\bf B#1} (#2) #3}
\newc\prep[3] 
                {{\em Phys. Rept.} {\bf #1} (#2) #3}
\newc\prl[3] 
                {{\em Phys. Rev. Lett.} {\bf #1} (#2) #3}
\newc\prd[3] 
                {{\em Phys. Rev.} {\bf D#1} (#2) #3}
\newc\ptp[3] 
                {{\em Prog. Theor. Phys.} {\bf #1} (#2) #3}
\newc\ppnp[3] 
                {{\em Prog. Part. Nucl. Phys.} {\bf #1} (#2) #3}
\newc\rmp[3] 
                {{\em Rev. Mod. Phys.} {\bf #1} (#2) #3 }
\newc\rpp[3] 
                {{\em Rept. Prog. Phys.} {\bf #1} (#2) #3 }
\newc\science[3] 
                {{\em Science} {\bf #1} (#2) #3 }
\newc\zpc[3] 
                {{\em Z. Phys.} {\bf C#1} (#2) #3}
\newc\err[3] 
           {{\em Erratum-ibid.} {\bf #1} (#2) #3 }

\newc\hepph[1] 
           {hep-ph#1 }

\newcount\hour
\newcount\minute
\newtoks\amorpm
\hour=\time\divide\hour by60 \minute=\time{\multiply\hour by60
\global\advance\minute by- \hour}
\edef\standardtime{{\ifnum\hour<12 \global\amorpm={am}%
   \else\global\amorpm={pm}\advance\hour by-12 \fi
   \ifnum\hour=0 \hour=12 \fi
   \number\hour:\ifnum\minute<100\fi\number\minute\the\amorpm}}
\edef\militarytime{\number\hour:\ifnum\minute<100\fi\number\minute}
\def\bold#1{\setbox0=\hbox{$#1$}%
    \kern-.025em\copy0\kern-\wd0
    \kern.05em\copy0\kern-\wd0
    \kern-.025em\raise.0433em\box0 }

\newcommand{\interv}[4]{$[#1,#2]$, $([#3,#4])$}

\newc\eg{{\rm {e.g.}}}  \newc\etal{{\rm {et al.}}} \newc\ie{{\rm i.e.}}
\newc\etc{{\rm {etc}}}
\newcommand\lsim{\mathrel{\rlap{\lower4pt\hbox{\hskip1pt$\sim$}}
   \raise1pt\hbox{$<$}}}
\newcommand\gsim{\mathrel{\rlap{\lower4pt\hbox{\hskip1pt$\sim$}}
   \raise1pt\hbox{$>$}}}
\newc{\mhalf}{m_{1/2}}      \newc{\mzero}{m_0}
\newc{\tanb}{\tan\beta}
\newc{\azero}{A_0}
\newc{\at}{A_t} \newc{\ab}{A_b} \newc{\atau}{A_\tau}
\newc{\bmu}{B\mu}           \newc{\sgn}{{\rm sgn}}
\newc{\mone}{M_1}           \newc{\mtwo}{M_2}

\newc{\charone}{\chi_1^\pm} \newc{\mcharone}{m_{\chi_1^\pm}}

\newc{\hl}{h}               \newc{\mhl}{m_{\hl}}   \newc{\gammahl}{\Gamma_{\hl}}
\newc{\hh}{H}               \newc{\mhh}{m_{\hh}}   \newc{\gammahh}{\Gamma_{\hh}}
\newc{\ha}{A}               \newc{\mha}{m_{\ha}}   \newc{\gammaha}{\Gamma_{\ha}}
\newc{\hpm}{H^{\pm}}        \newc{\mhpm}{m_{\hpm}} \newc{\gammahpm}{\Gamma_{\hpm}}
\newc{\hp}{H^{+}} \newc{\mhp}{m_{\hp}} \newc{\hm}{H^{-}}
\newc{\mhm}{m_{\hm}}
\newc{\xt}{X_{t}}           \newc{\xb}{X_{b}}

\newc{\qzero}{Q_0}          \newc{\qstop}{Q_{\widetilde t}}
\newc{\amu}{a_{\mu}}        \newc{\amususy}{a_{\mu}^{\text{SUSY}}}
\newc{\amuexpt}{a_{\mu}^{\text{expt}}}        \newc{\amusm}{a_{\mu}^{\text{SM}}}
\newc{\deltaamususy}{\delta a_{\mu}^{\text{SUSY}}}

\newc\gmtwo{(g-2)_{\mu}} \newc\deltaamu{\Delta a_{\mu}}
\newc{\msbar}{\overline{MS}} \newc{\drbar}{\overline{DR}}
\newc{\yt}{h_t} \newc{\yb}{h_b} \newc{\ytau}{h_{\tau}}

\newc{\mtop}{m_t}               \newc{\mtpole}{M_t}
\newc{\mtaupole}{m_{\tau}^{\text{pole}}}
\newc{\mtmtsmmsbar}{m_t(m_t)^{\msbar}_{{\text{SM}}}}
\newc{\mtmtsmdrbar}{m_t(m_t)^{\drbar}_{{\text{SM}}}}
\newc{\mtmtmssmdrbar}{m_t(m_t)^{\drbar}_{{\text{SUSY}}}}

\newc{\mbmbmsbar}{m_b(m_b)^{\msbar} }

\newc{\mbmbsmmsbar}{m_b(m_b)^{\msbar}_{{\text{SM}}}}
\newc{\mbmzsmmsbar}{m_b(\mz)^{\msbar}_{{\text{SM}}}}
\newc{\mbmzsmdrbar}{m_b(\mz)^{\drbar}_{{\text{SM}}}}
\newc{\mbmzmssmdrbar}{m_b(\mz)^{\drbar}_{{\text{SUSY}}}}

\newc{\mtaumzsmmsbar}{m_{\tau}(\mz)^{\msbar}_{{\text{SM}}}}
\newc{\mtaumzsmdrbar}{m_{\tau}(\mz)^{\drbar}_{{\text{SM}}}}
\newc{\mtaumzmssmdrbar}{m_{\tau}(\mz)^{\drbar}_{{\text{SUSY}}}}

\newc{\mgut}{M_{\rm GUT}}
\newc{\mplanck}{M_{\rm P}}      \newc{\mpl}{M_{\text{Pl}}}
\newc{\msusy}{M_{\rm SUSY}}      \newc{\ms}{M_{\text{S}}}
\newc{\jxf}{J({\xf})}
\newc{\jxfexact}{J_{\rm exact}({\xf})}  \newc{\jxfexp}{J_{\rm exp}({\xf})}
\newc{\VEV}[1]{\langle #1 \rangle}
\newc{\xf}{x_f}
\newc\vrel{v_{\rm rel}}

\newcommand\mchi{m_{\chi}}              

\newc\sell{{\widetilde e}_L}      \newc\msell{m_{\sell}}
\newc\selr{{\widetilde e}_R}      \newc\mselr{m_{\selr}}
\newc\snue{{\widetilde \nu}_e}      \newc\msnue{m_{\snue}}
\newc\snutau{{\widetilde \nu}_\tau}      \newc\msnutau{m_{\snutau}}
\newc\supl{{\widetilde u}_L}      \newc\msupl{m_{\supl}}
\newc\supr{{\widetilde u}_R}      \newc\msupr{m_{\supr}}
\newc\sdl{{\widetilde d}_L}      \newc\msdl{m_{\sdl}}
\newc\sdr{{\widetilde d}_R}      \newc\msdr{m_{\sdr}}

\newc\squark{\widetilde q} \newc\squarkl{{\widetilde q}_L} \newc\squarkr{{\widetilde q}_R}
\newc\msquark{m_{\squark}}
\newc\msquarkl{m_{\squarkl}} 
\newc\msquarkr{m_{\squarkr}}

\newcommand\slepton{\widetilde l} 
   \newcommand\mslepton{m_{\slepton}}
\newcommand{\msl}{\mslepton}
\newcommand{\msq}{\msquark}

\newcommand\stauone{{\widetilde \tau}_1}   
\newcommand\stautwo{{\widetilde \tau}_2}   

\newcommand\gluino{\widetilde g}

\newc\sfermion{\tilde f}  \newc\msfermion{m_{\sfermion}}
\newc\cmeter{{\rm cm}} \newc\meter{{\rm m}} \newc\kmeter{{\rm km}}
\newc\second{{\rm sec}}
\newc{\gstar}{g_\ast}           \newc{\gsstar}{g_{s\ast}}
\newc{\geff}{g_{\rm eff}}
\newcommand\mz{m_{Z}}

\newc{\sthw}{\sin\theta_W}              \newc{\cthw}{\cos\theta_W}
\newc{\bino}{\widetilde B}              \newc{\wino}{\widetilde W_30}
\newc{\higgsinob}{{\widetilde H}^0_b}   \newc{\higgsinot}{{\widetilde H}^0_t}
\newc{\abund}{\Omega h^2}
\newc{\abundchi}{\Omega_\chi h^2}
\newc{\abundcdm}{\Omega_{\text{CDM}} h^2}
\newc{\omcd}{\abundcdm}
\newc{\omegam}{\Omega_{M}}       \newc{\abundm}{\Omega_{M} h^2}
\newc{\omegab}{\Omega_{b}}       \newc{\abundb}{\Omega_{b} h^2}
\newc{\omegacdm}{\Omega_{CDM}}
\newc{\omegatot}{\Omega_{TOT}}
\newc{\rhocrit}{\rho_{crit}}
\newc{\rhochi}{\rho_{\chi}}
 \newcommand\fb{\,\mbox{fb}}

\newc\BR{BR}
\newc\bsgamma{b\rightarrow s \gamma }
\newc\bxsgamma{\overline{B}\rightarrow X_{s}\gamma}
\newc\brbsgamma{\BR(\overline{B}\rightarrow X_s\gamma)}

\newc{\beq}{\begin{equation}}
\newc{\eeq}{\end{equation}}

\newc\neutone{\chi_1^0} \newc\mneutone{m_{\neutone}}
\newc\neuttwo{\chi_2^0} \newc\mneuttwo{m_{\neuttwo}}
\newc\neutthree{\chi_3^0} \newc\mneutthree{m_{\neutthree}}
\newc\neutfour{\chi_4^0} \newc\mneutfour{m_{\neutfour}}

\newc\stoponetwo{{\widetilde t}_{1,2}}
\newc\sbotonetwo{{\widetilde b}_{1,2}}
\newc\stauonetwo{{\widetilde \tau}_{1,2}}

\newc{\sigsip}{\sigma^{SI}_{p}} \newc{\sigsin}{\sigma^{SI}_{n}}
\newc{\sigsiN}{\sigma^{SI}_{N}}
\newc{\sigsdp}{\sigma^{SD}_{p}} \newc{\sigsdn}{\sigma^{SD}_{n}}
\newc{\sigsiA}{\sigma^{SI}_{A}}

\newc\xilim{\xi_{\rm lim}} 
\newc\tlim{t_{\rm lim}} 
\newc\zetalim{\zeta_{\rm lim}} 

\newc\zetah{\zeta_h}
\newc{\relprobone}[1]{p({#1} \vert d)}
\newc{\relprobtwo}[2]{p({#1},{#2} \vert d)}

\long\def\begincomment#1\endcomment{%
       \begingroup\sf\baselineskip12pt#1\endgroup}

\newcommand{\squishlist}{
  \begin{list}{$\bullet$}
   { \setlength{\itemsep}{0pt}      \setlength{\parsep}{3pt}
     \setlength{\topsep}{3pt}       \setlength{\partopsep}{0pt}
     \setlength{\leftmargin}{1.em} \setlength{\labelwidth}{1em}
     \setlength{\labelsep}{0.5em} } }
\newcommand{\squishend}{
   \end{list}  }
        
\newcommand{\data}{d}

\newcommand{\basis}{m}

\newcommand{\prof}{{\mathfrak L}}

\newcommand{\chisq}{\chi^2}

\newc{\ww}{0.49\linewidth}
\newc{\ttr}{0.32\linewidth} 
\newc{\qq}{0.24\linewidth}

\newcommand{\thstar}{\theta_\star}
\newcommand{\thML}{\theta_{\text{ML}}}

\begin{document}

\begin{titlepage}
\pagestyle{empty}
\baselineskip=21pt
\vskip 1.5cm
\begin{center}

{\Large\bf Efficient reconstruction of CMSSM parameters from LHC data -- A case study}

\end{center}   
\begin{center}   
\vskip 0.75 cm
{\bf Leszek Roszkowski}${}^a$\,\footnote{\, L.Roszkowski@sheffield.ac.uk}, 
{\bf Roberto Ruiz de Austri}${}^b$\,\footnote{\, rruiz@ific.uv.es},
and {\bf Roberto Trotta}${}^c$\,\footnote{\, r.trotta@imperial.ac.uk}
\vskip 0.1in
\vskip 0.4cm ${}^a$ {\it Department of Physics and
Astronomy, University of Sheffield, Sheffield, S3 7RH, UK, and\\ The Andrzej
Soltan Institute for Nuclear Studies, Warsaw, Poland}\\ 
${}^b$ {\it Instituto de F\'isica Corpuscular, IFIC-UV/CSIC, Valencia, Spain}\\ 
${}^c$ {\it Astrophysics Group, Imperial College, 
       Blackett Laboratory, Prince Consort Road, London SW7 2AZ, UK}

\vskip 1cm
\abstract{We present an efficient method of reconstructing the
  parameters of the Constrained MSSM from assumed future LHC data,
  applied both on their own right and in combination with the
  cosmological determination of the relic dark matter abundance.
  Focusing on the ATLAS SU3 benchmark point, we demonstrate that our
  simple Gaussian approximation can recover the values of its
  parameters remarkably well. We examine two popular non-informative
  priors and obtain very similar results, although when we use an
  informative, naturalness-motivated prior, we find some sizeable
  differences.  We show that a further strong improvement in
  reconstructing the SU3 parameters can by achieved by applying
  additional information about the relic abundance at the level of
  WMAP accuracy, although the expected data from Planck will have only
  a very limited additional impact. Further external data may be
  required to break some remaining degeneracies.  We argue that the
  method presented here is applicable to a wide class of low-energy
  effective supersymmetric models, as it does not require to deal with
  purely experimental issues, \eg, detector performance, and has the
  additional advantages of computational efficiency. Furthermore, our
  approach allows one to distinguish the effect of the model's
  internal structure and of the external data on the final parameters
  constraints.  }
\end{center}
\baselineskip=18pt \noindent

\vfill
\end{titlepage}


\section{Introduction}\label{sec:intro}

If softly broken low-energy supersymmetry (SUSY) provides a correct
description of the particle physics realm at energy scales around a
few hundred GeV and above, then superpartners are likely to be
discovered at the LHC. One of the main goals of ATLAS and CMS
experiments will be to identify those particles by determining their masses and other properties.

The actual outcome will depend not only on the LHC machine and
detector performance but obviously also on the mass scales of the
superpartners themselves. A whole plethora of different possibilities
can be listed here, ranging from one extreme where all of the
superpartners may come out to be too heavy to for the LHC reach, to
another where all, or most, of them will be discovered. Unfortunately,
basically the whole spectrum of options remains open even in perhaps
the most economical SUSY framework, the Constrained Minimal
Supersymmetric Model (Constrained MSSM, or CMSSM)~\cite{kkrw94} which
includes the minimal supergravity (mSUGRA) model~\cite{sugra-refs}, as
shown by a number of recent global fits of the CMSSM based on Bayesian
statistics~\cite{allanach-bayes,rrtetal} 
and on a $\chi^2$ approach~\cite{ehow06,buchmueller08}. While the
latter show a stronger preference for a fairly low SUSY mass scale, in
the range of a few hundred GeV, the former point to a more cautious
picture, where a much wider mass range remains allowed. This
discrepancy is caused by the fact that, with the data that is
currently available, even the CMSSM still remains to some extent
underconstrained, and the specifics of the statistical and data
analysis treatment can lead to fairly different results.  It is
therefore clear that selecting, or at least limiting, SUSY models by
using LHC measurements is certainly going to be a very challenging
task as there exist large degeneracies among the MSSM parameters that
can lead to indistinguishable LHC signatures (see, \eg,
Ref.~\cite{ArkaniHamed:2005px}).

In preparation for dealing with real data, a number of approaches to particle mass reconstruction have
been developed based on extracting
kinematic information from one or more decay chains of superpartners,
typically requiring two or more visible particles in the final
state~\cite{massreconstruction-refs}. These and other techniques have
been used by LHC experimental groups which have performed a large
number of detailed studies in a few reference, or ``benchmark'',
points, often selected in such a way as to typically allow several
of the superpartners to be seen at the LHC.

In a recent extensive ATLAS Report~\cite{atlas09}, in the framework of
the CMSSM/mSUGRA a so-called ATLAS SU3 benchmark point (which is
specified below) was examined with Markov Chain Monte Carlo (MCMC)
scans, with the aim of evaluating the expected accuracy of
reconstructing CMSSM parameters: a common gaugino mass parameter
$\mhalf$, a common scalar mass parameter $\mzero$, a common trilinear
term $\azero$, all evaluated at the unification scale $\mgut\simeq
2\times10^{16}\gev$, plus a Higgs vacuum expectation values $\tanb$.
Assuming an integrated luminosity of $1\fb^{-1}$, a dilepton and
lepton+jets edge analysis of the decay chain $\squarkl \rightarrow
\chi^0_2 (\rightarrow \slepton^{\pm} l^{\mp}) q \rightarrow\chi^0_1
l^+ l^- q$ and the high-$p_T$ and large missing energy analysis of the
decay chain $\squarkr \rightarrow \chi^0_1 q$ were performed, where
$\squarkl$ ($\squarkr$) denotes the first or second generation left
(right) squark, $\chi^0_{1,2}$ the first and second neutralino and
$\slepton$ an intermediate slepton. It was concluded that $\mhalf$ and
$\mzero$ could be reconstructed with adequate accuracy, while prospects
for $\tanb$ looked somewhat poorer, and even more so for
$\azero$~(see Ref.~\cite{atlas09}, pp.~1617, ff).

In this paper we perform an independent analysis of the ATLAS SU3
point using the publicly available information about the expected
ATLAS capabilities to measure the SU3 mass spectrum. We first
demonstrate that a simple modelling of the mass spectrum constraints
in an effective likelihood is sufficient to reproduce with reasonable
accuracy the results of the full ATLAS analysis, while being much more
economical in terms of computational requirements, when we use the
same linear, or flat (as defined below) prior.  We then build on the
ATLAS analysis by examining the impact of two other priors. We find
that non-informative priors (\ie, priors whose characteristic scale is
much larger than the support of the likelihood) lead to approximate
prior-independence in the posterior, thus significantly improving with
respect to the current situation.  On the other hand, if one imposes
extra theoretical prejudice in the prior (by choosing a prior that
penalizes fine-tuning), the posterior is still quite strongly
affected.

We also
compare with the limits that can be obtained using a maximum
likelihood analysis, and we show that the choice of statistics
(Bayesian or maximum likelihood) no longer matters once one combines
ATLAS data with cosmological relic abundance determinations. We
clarify what r\^{o}le is played by assuming a specific theoretical model
(here the CMSSM) in complementing the information coming from the
ATLAS measurements with model-specific theoretical correlations
between masses of the observables. 

In the second step, we go beyond the ATLAS analysis by applying
additional information about the cosmological relic density
$\abundchi$ of the lighest neutralino $\neutone$ (below often denoted
by $\chi$ for simplicity), assumed to be dark matter (DM) in the
Universe. The dark matter abundance clearly provides
additional information about the model at hand, and in this
analysis we aim at obtaining a quantitative measure for the extra
constraining power that it provides, on top of that expected from the
ATLAS data.  Here we first impose WMAP uncertainties on $\abundchi$
and demonstrate a significant improvement in the determination of the
CMSSM parameters, especially $\mzero$. Next we investigate the impact
of further reducing the observational errors on $\abundchi$ to an accuracy as
expected from Planck, and show that this will lead to only a further modest
improvement. We also comment on the impact of some other commonly used
constraints, in particular from $\bsgamma$ and $\gmtwo$, which we,
however, do not apply here. Finally, we examine the impact of
the different constraints from ATLAS and cosmology on the
uncertainties of mass measurements of several superpartners and on
predictions for the scattering cross section relevant for direct
detection of dark matter experiments.

The paper is organized as follows. In Sec.~\ref{setup:sec} we present
a setup of our analysis for the ATLAS SU3 point in the CMSSM,
including details of our scans of the CMSSM parameters.  In
Sec.~\ref{sec:results} we present our numerical results for the
posterior probability density functions (pdfs), including a discussion
of an impact of adding further assumptions and information. In Sec.~\ref{sec:pl}
we compare some of these with the alternative measure of the profile
likelihood. We summarize our findings in Sec.~\ref{sec:summary}.

\section{Setup and benchmark point}\label{setup:sec}

\subsection{The ATLAS SU3 benchmark point} \label{sec:su3}

We examine the ATLAS SU3 benchmark point for which input values of
CMSSM parameters are given on the left side of
Table~\ref{tab:su3input}. Since in the ATLAS analysis errors of
relevant SM parameters (``nuisance parameters'') were not included, we
assume that the benchmark values for the nuisance parameters are set
at their central values as given on the right side of
Table~\ref{tab:su3input}. In the reconstruction done below, we then
allow the nuisance parameters to vary and we constrain them using the
likelihood given below.

\begin{table}[t!hb]
\centering   

\begin{tabular}{| l l || l l | }
\hline
CMSSM parameter  &  ATLAS SU3 benchmark value & SM parameter & Input value  \\\hline
$\mhalf$ & 300 GeV & $\mtpole$  &  172.6 GeV \\
$\mzero$ & 100 GeV & $\mbmbmsbar$ &4.20 GeV   \\ 
$\tanb$ & 6.0 & $\alphas$       &   0.1176  \\
$\azero$ & $-300$ GeV & $1/\alphaemmz$  & 127.955 \\
$\sgn(\mu)$ & $+$ & & \phantom{a}\\
\hline 
\end{tabular}
\caption{Left side: input CMSSM parameters values for the ATLAS SU3
  benchmark point. Right side: input values of relevant SM
  parameters used in the numerical analysis. }
\label{tab:su3input}
\end{table}

\begin{table}[b!ht]
\centering   
\begin{tabular}{| l l | l l | l l | }
\hline
superpartner & mass & superpartner & mass & superpartner &mass \\ 
\hline \hline
$\neutone(=\chi)$ & 117.9 GeV & $\widetilde{e}_L$, $\widetilde{\mu}_L$ & 230.8 GeV & $\widetilde{d}_L$ & 666.2 GeV \\
$\neuttwo$ & 223.4 GeV & $\widetilde{e}_R$, $\widetilde{\mu}_R$ & 157.5 GeV & $\widetilde{d}_R$ & 639.0 GeV\\
$\chi_3^0$ & 463.8 GeV & $\widetilde{\nu}_e$, $\widetilde{\nu}_\mu$ & 217.5 GeV & $\widetilde{u}_L$ & 660.3 GeV \\
$\chi_4^0$ & 479.9 GeV & $\widetilde{\tau}_1$ & 152.2 GeV & $\widetilde{u}_R$ & 644.3 GeV\\
$\chi_1^+$ & 224.4 GeV & $\widetilde{\tau}_2$ & 232.4 GeV & $\widetilde{b}_1$ & 599.0 GeV \\
$\chi_2^+$ & 476.4 GeV & $\widetilde{\nu}_\tau$ & 216.9 GeV & $\widetilde{b}_2$ & 636.6 GeV \\
$\widetilde{g}$ & 717.5 GeV & & & $\widetilde{t}_1$ & 446.9 GeV \\
     & & & & $\widetilde{t}_2$ & 670.9 GeV \\
\hline
\end{tabular}
\caption{Superpartner mass spectrum for the ATLAS SU3 point. \label{tab:su3spectra}} 
\end{table}

Since LHC data is rather unlikely to differentiate among the flavors
of the squarks of the first two generation, in what follows we denote
them all by a common symbol $\squark$, and by $\msq$ their average
mass, similarly as in Ref.~\cite{atlas09}. On the other hand,
$\slepton$ will denote the lightest slepton and $\msl$ its mass. In
the case of the ATLAS SU3 point its r\^{o}le is played by
$\widetilde{\tau}_1$.

The resulting mass spectrum, as computed using the SoftSusy code 
version 1.0.18~\cite{softsusy} in
the 1-loop approximation is given in
Table~\ref{tab:su3spectra}. By comparing with the mass spectrum for
the ATLAS SU3 point given in Ref.~\cite{atlas09} (cf.~Table~2 on page
1516), we can see some differences, especially a systematic shift in
squark masses by a few tens of GeV, which may be due to using
different numerical codes, approximations (although in both cases 1
loop expressions are applied) as well as different inputs in SM
parameter values. While this will contribute to some differencies we
will find with the ATLAS results, at the end these discrepancies are
of secondary importance, as we discuss below.

\subsection{The likelihood function}\label{sec:likefn}
The study performed by the ATLAS Collaboration on the SU3 point reports the 
expected
accuracy in the reconstruction of some of the masses and mass
differences in the benchmark SUSY spectrum given in
Table~\ref{tab:su3spectra}. Dilepton edges will constrain
$m_{\chi_1^0}$, $m_{\chi_2^0}$, $\msq$ and $\msl$ with fairly poor
accuracy, while providing much tighter limits on the mass
differences between the three latter quantities and the lightest
neutralino, since these follow more directly from endpoint
measurements. 
The endpoint in the dilepton invariant mass
distribution is determined by the masses of the particles involved. In
the case of the SU3 point considered here, the two-body decay channel
$\chi^0_2 \rightarrow \slepton^{\pm} l^{\mp}$ dominates, since
$\mneuttwo>\mslepton$, and the distribution of the invariant mass of
the two leptons is triangular, with an endpoint given
by~\cite{atlas09} (p.~1619)
\be
m_{}^{\rm edge} =
\mneuttwo\sqrt{1-\left(\frac{\mslepton}{\mneuttwo}\right)^2 }
\sqrt{1-\left(\frac{\mneutone}{\mslepton}\right)^2 }.
\label{eq:leptonedge}
\ee
A measurement of the dilepton endpoint leads to a relationship between 
$\neuttwo$, $\neutone$ and the slepton involved. Further mass
distributions  are considered in order to
determine the masses of all the particles involved in the process,  as
described in Ref.~\cite{atlas09} (pp.~1619, ff), along with event and
cut selection procedures adopted in a reconstruction of the dilepton
and other edges.

The observable quantities to be constrained by ATLAS are
given by the set
\be
\theta = \{ \mneutone, \mneuttwo-\mneutone, \msl-\mneutone,
\msq-\mneutone \}.
\label{eq:thetavalues}
\ee
We further assume that the maximum likelihood (ML) value of $\theta$
obtained by ATLAS, $\thML$, corresponds to the value of the true
benchmark point, $\thstar = \{117.9, 105.5, 34.3, 534.5\}\gev$, where
the numerical value is obtained from Table~\ref{tab:su3spectra}. In
other words, we neglect realization noise, an assumption which is
justified by the fact that $\langle \thML \rangle = \thstar$, where
$\langle \cdot \rangle$ denotes an average over realizations. The
likelihood function from ATLAS is then modeled as a Gaussian centered
around the true value of the observable quantities,
\be \label{eq:likeATLAS}
-2 \ln \like_\text{ATLAS} = \chi^2_\text{ATLAS} = (\theta - \thML)^t C^{-1} (\theta - \thML),
\ee
where the covariance matrix $C$ is given in Table~\ref{tab:covmat} (and we have dropped an irrelevant 
normalization constant). It
represents the full covariance between the masses and the mass
differences. The covariance matrix includes statistical
errors only; systematic errors are negligible.

The form of the ATLAS likelihood function given in
Eq.~\eqref{eq:likeATLAS} is a simple Gaussian approximation to the
actual likelihood function that one would obtain from a full analysis
of simulated ATLAS data. The latter is, however, not available outside
the Collaboration, and therefore our approximation represents the best
that can be reasonably done given the information that is expected to be publicly
available.
There are two reasons why it might be interesting to consider an
approximate ATLAS likelihood function at the level of the SUSY mass
spectrum. Firstly, it is not unreasonable that the simple
approximation adopted here will give a fairly accurate representation
of ATLAS capabilities (see below for further comments on how this
compares with the full analysis carried out by the ATLAS
Collaboration), and therefore provide a useful shortcut to
quantitatively implementing the constraints that would result from a full
analysis. Secondly, given the model-independent constraints on the
low-energy SUSY mass spectrum, one can use them to constrain
the high-energy parameters of any SUSY model of the MSSM class, in
this case the CMSSM. As we 
shall show below, the constraining power included in the low-energy
likelihood is then supplemented by the theoretical structure of the
model itself.

\begin{table}
\centering    

\begin{tabular}{| l | l l l l |}
\hline
& $\mneutone$ &  $\mneuttwo-\mneutone$ & $\msl-\mneutone$ &$\msq-\mneutone$ \\ \hline 
$\mneutone$ & $3.72\times10^3$  & $53.40$ &  $1.92\times10^3$ &   $10.75\times10^2$ \\
$\mneuttwo-\mneutone$ &   &    $3.6$ &    $29.0$ &  $-1.3$ \\ 
$\msl-\mneutone$ &  & & $1.12\times10^3$  &  $4.65$ \\
$\msq-\mneutone$ &  & & & $14.1$\\\hline
\end{tabular}
\caption{ATLAS covariance matrix employed in the analysis. 
\label{tab:covmat}}
\end{table}


We will now examine prospects for reconstructing the input values
of the ATLAS SU3 point. We
consider the following data combinations:
\begin{itemize}
\item  ATLAS data only: including only the likelihood function given
  by Eq.~\eqref{eq:likeATLAS};
\item  ATLAS data (as above) + WMAP-level uncertainty on dark matter abundance;
\item  ATLAS data (as above) + Planck-level uncertainty on dark matter abundance.
\end{itemize}

When including information about the cosmological dark matter
abundance, we have to address the fact that the value of the
neutralino relic abundance (computed using the code
Micromegas~\cite{micro}) for the ATLAS SU3 benchmark point, is
$\abundchi = 0.2332$, which is a factor of some 2.5 above the WMAP
range of $0.1099 \pm 0.0062$ for the cosmological dark
matter~\cite{wmap5yr}. However, assuming standard Big Bang cosmology
and that the CMSSM is correct, we would expect that for the actual
measured value of the CMSSM parameters, for which ATLAS and/or CMS
measurements are made, $\abundchi$ would lie in the WMAP range.  It is
therefore not unreasonable to examine the impact of the extra piece of
information, both in terms of its value and uncertainty, which is
provided by the cosmological relic density of dark matter. Notice that
we do not rely here on a potential ability to reconstruct $\abundchi$
from LHC data alone~\cite{Nojiri:2005ph} (which, on the other hand,
could provide an important cross-check of our cosmological input) but
use it as external constraint. Therefore, since the ATLAS analysis has
been carried out for the ATLAS SU3 point, with the specific values of
the CMSSM parameters, we also adopt to keep this central value for the
relic abundance. On the other hand, we don't expect any major change
in the numerical results presented below if $\abundchi$ were actually
close to the WMAP range.

In our analysis we implement the
cosmological dark matter determination as follows. Firstly, we assume that the true
value of the relic abundance corresponds to the value computed for  the benchmark
point, and that WMAP-level constraints correspond to a Gaussian
likelihood centered around that value with standard deviation given by
the current WMAP uncertainty, namely $\sigma_\text{WMAP} =
6.2\times10^{-3}$: 
\be \label{eq:likeWMAP} 
-2 \ln \like_\text{WMAP} = \chi^2_\text{WMAP} =
\frac{(\abundchi - 0.2332)^2}{\sigma_\text{WMAP}^2}. 
\ee 
The Planck satellite is expected to improve the WMAP accuracy on the
relic abundance by a factor $\sim 10$~\cite{planckdm}.  When
including Planck-level constraints, we therefore adopt the same
likelihood as above but with a smaller standard deviation,
$\sigma_\text{Planck} = 6.2\times10^{-4}$.  Since for the ``bulk
region'' to which the ATLAS SU3 point belongs the theoretical error in
the relic abundance is estimated to be tiny~\cite{Allanach:2004jh} we neglect
it here.  In other cases it can be much larger, primarily due to the
larger uncertainties in computing mass spectra; for example in the
focus point region it would likely dominate and this would cloud
the potential impact of the cosmological data.

In order to facilitate a comparison with the ATLAS
study~\cite{atlas09}, in this analysis we do not apply any other
constraints, \eg, from LEP, rare processes in heavy quark physics
($\bsgamma$, \etc) or the anomalous magnetic moment of the muon
$\gmtwo$, which are routinely used in global analyses of the CMSSM and
other popular SUSY models.  We have also checked that fixing the
nuisance parameters or marginalising over them has a negligible impact
on the results. Therefore we only present results with the nuisance SM
parameters fully marginalised, even though in the ATLAS analysis the
SM parameters were fixed at their central values.

\subsection{Scanning the CMSSM parameters}\label{sec:cmssmps}

With the aim of reconstructing the true values of the defining parameters for
the ATLAS SU3 point, we explore the CMSSM parameter space with the
help of 
the Nested Sampling (NS)
scanning technique, as implemented in the MultiNest algorithm~\cite{NS}. 

We consider two different non-informative priors, that is, priors which contain minimal
assumptions about the values of the parameters: 
\begin{itemize}
\item {\bf flat prior:} flat in $\mhalf, \mzero, \azero, \tanb$,
  with the ranges: $50 \gev \leq \mhalf, \mzero \leq 500
  \gev$, $2 \leq \tanb \leq 62$ and $-4 \tev \leq \azero\leq 4 \tev$. 

\item {\bf log prior:} flat in $\log\mhalf, \log\mzero, \azero,
  \tanb$, with the same ranges: $\log(50) \leq \log\mzero, \log\mhalf
  \leq \log(500) $ (in GeV), and as above for $\azero$ and $\tanb$. 
 \end{itemize}  

 Notice that we have employed here narrower ranges of $\mhalf$ and
 $\mzero$ than the values of up to a few TeV used in our previous
 analyses~\cite{rrtetal, tfhrr1}. However, we have checked that
 enlarging the prior range to much larger values of $\mhalf, \mzero$
 (up to 4 TeV) has no impact on our reconstructed parameter values, as
 our algorithm correctly recovers the true parameter values even in
 the case of a much larger prior range.  Finally, for the SM nuisance
 parameters we assume the same ranges as in our previous
 papers~\cite{rrtetal, tfhrr1}; in any case as we have mentioned
 above, the details of the treatment of nuisance parameters has
 basically no impact on the results presented here.

 One of the aims of this work is to demonstrate that ATLAS data will
 achieve approximate prior-independence for the two choices of
 non-informative priors given above, which have been widely used in
 the literature so far and for which it has been shown that the
 current posterior for the CMSSM retains a fairly substantial prior
 dependence~\cite{tfhrr1}.  However, while this is encouraging,
 clearly that does not imply that one should expect the same to hold
 with any other choice of prior, of which there is an infinite
 range. Furthermore, in the Bayesian framework it is always possible
 to supplement the information contained in the likelihood by external
 prior information, for example by imposing ``naturalness''
 constraints~\cite{ccr08,bclw07}. In this case, one would not expect
 the posterior to remain independent of the prior, but actually to
 show stronger constraints than for the case of non-informative
 priors. In order to investigate to what extent a ``naturalness''
 prior can supplement ATLAS data in constraining CMSSM parameters, we
 also consider an informative prior choice in
 section~\ref{sec:naturalness}, with the following ``CCR prior''
 (after Cabrera, Casas and Ruiz de Austri, who introduced
 it)~\cite{ccr08},
\begin{itemize}
\item {\bf CCR prior:} flat on $\mzero, \mhalf, \azero, B$
  but with an effective  ``penalty term'' that naturally leads to low
  fine tuning among SUSY parameters.
 \end{itemize}  

The CMSSM is often treated as an effective theory following from
mSUGRA which is parametrized in terms of the following 
parameters  $\{\mzero, \mhalf, \azero, B, \mu, \}$, which are then
treated as in some sense more fundamental. 
On the other hand, for the purpose of performing a numerical scaning  of 
the model parameter space it is much more convenient to trade some of them for
the CMSSM parameters which have a more direct phenomenological
significance. (In addition one has the usual SM nuisance parameters.)
In Refs.~\cite{ccr08, Cabrera:2009dm} it has been shown that it is
convenient to replace $\mu$ with $\mz$, which is trivially integrated out. 
In addition, this procedure automatically takes into accounts 
the usual measure~\cite{Ellis:1986yg, Barbieri:1987fn} of the 
degree of fine-tuning. Furthermore, it is also convenient to trade
the $B$ parameter for the derived quantity $\tanb$. 

The change of variables $\{\mu,B\}\ \rightarrow\
\mz,\tanb$ leads to the {\em effective prior} in the CMSSM variables
\bea
\label{eff_prior}
p_{\rm eff}(\mzero, \mhalf, \azero, \tan\beta)\  \equiv\ 
J|_{\mu=\mu_Z}\  p(\mzero, \mhalf, \azero, B, \mu=\mu_Z)\, ,
\eea
where $p(\mzero, \mhalf, \azero, B, \mu=\mu_Z)$ the prior for the
mSUGRA parameters and $J$ is a Jacobian of the transformation, which
is evaluated in the usual way from minimization equations of the Higgs
scalar potential, $V(H_1, H_2)$. This way one arrives at an approximate form
for the effective prior~\cite{ccr08}
\bea
\label{eq:approx_eff_prior}
p_{\rm eff}(\mzero, \mhalf, \azero, \tanb)\ \ \propto  \
\frac{\tan^2\beta-1}{\tan^2\beta (1+\tan^2\beta)} \frac{B_{\rm low}}{\mu_Z}  
p(\mzero, \mhalf, \azero, B, \mu=\mu_Z)\ 
,
\eea
where $B_{\rm low}$ is the parameter $B$ evaluated at the electroweak
scale and $\mu_Z$ is chosen so that it gives the correct of $\mz$.
The CCR prior is then defined as the effective prior $p_{\rm
  eff}(\mzero, \mhalf, \azero,
\tanb)$,~Eq.~\eqref{eq:approx_eff_prior}, where we take a flat prior
in $\mzero, \mhalf, \azero, B$ and $\mu$.

\section{Results}\label{sec:results} 

In this Section we present our numerical results from scans performed
using the publicly available SuperBayes package version
1.35~\cite{superbayes}, which we have modified in order to include a
Gaussian likelihood from projected ATLAS data as described above.

\begin{figure}[tbh!]
\begin{center}
\includegraphics[width=\ww]{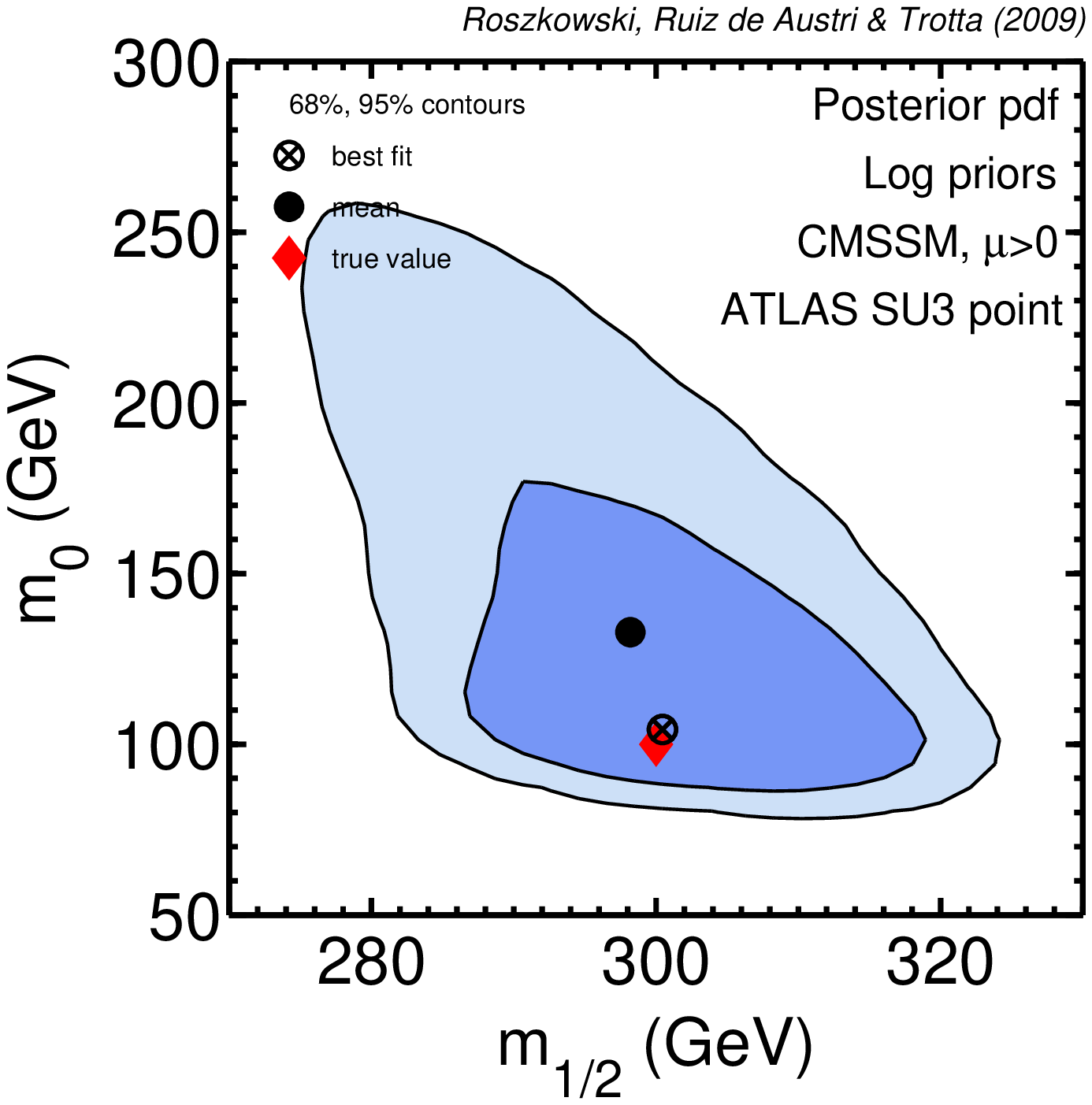}
\includegraphics[width=\ww]{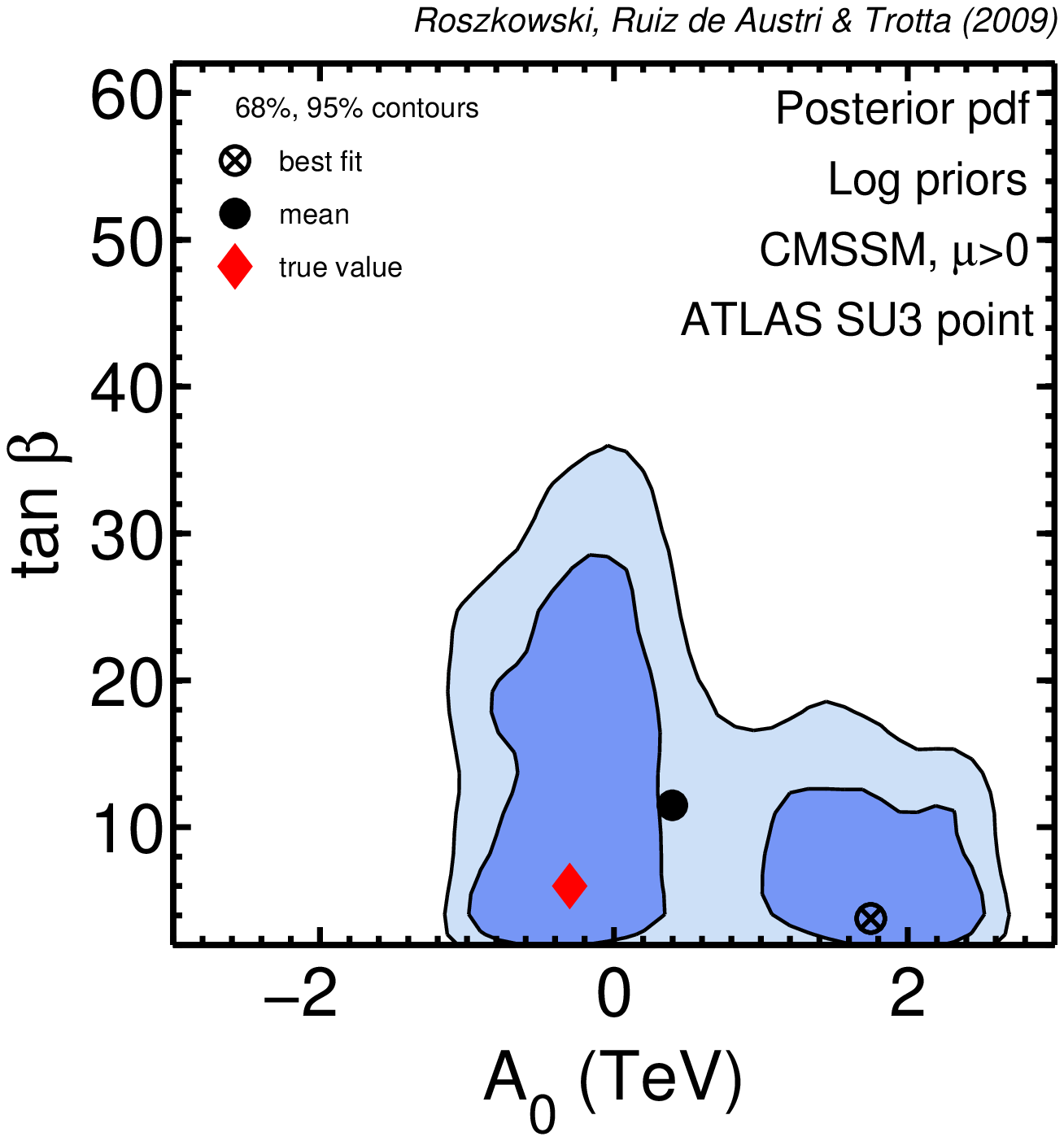} 
\caption[test]{2D posterior pdf for the case of applying ATLAS mass spectrum data alone,
  for some CMSSM parameter combinations and for the log prior
  choice. Compare with Fig. 12, p.~1638 
  of~\cite{atlas09}. }
\label{fig:CMSSM_LHC_2D_log}
\end{center}
\end{figure}

\begin{figure}[tbh!]
\begin{center}
\includegraphics[width=\ww]{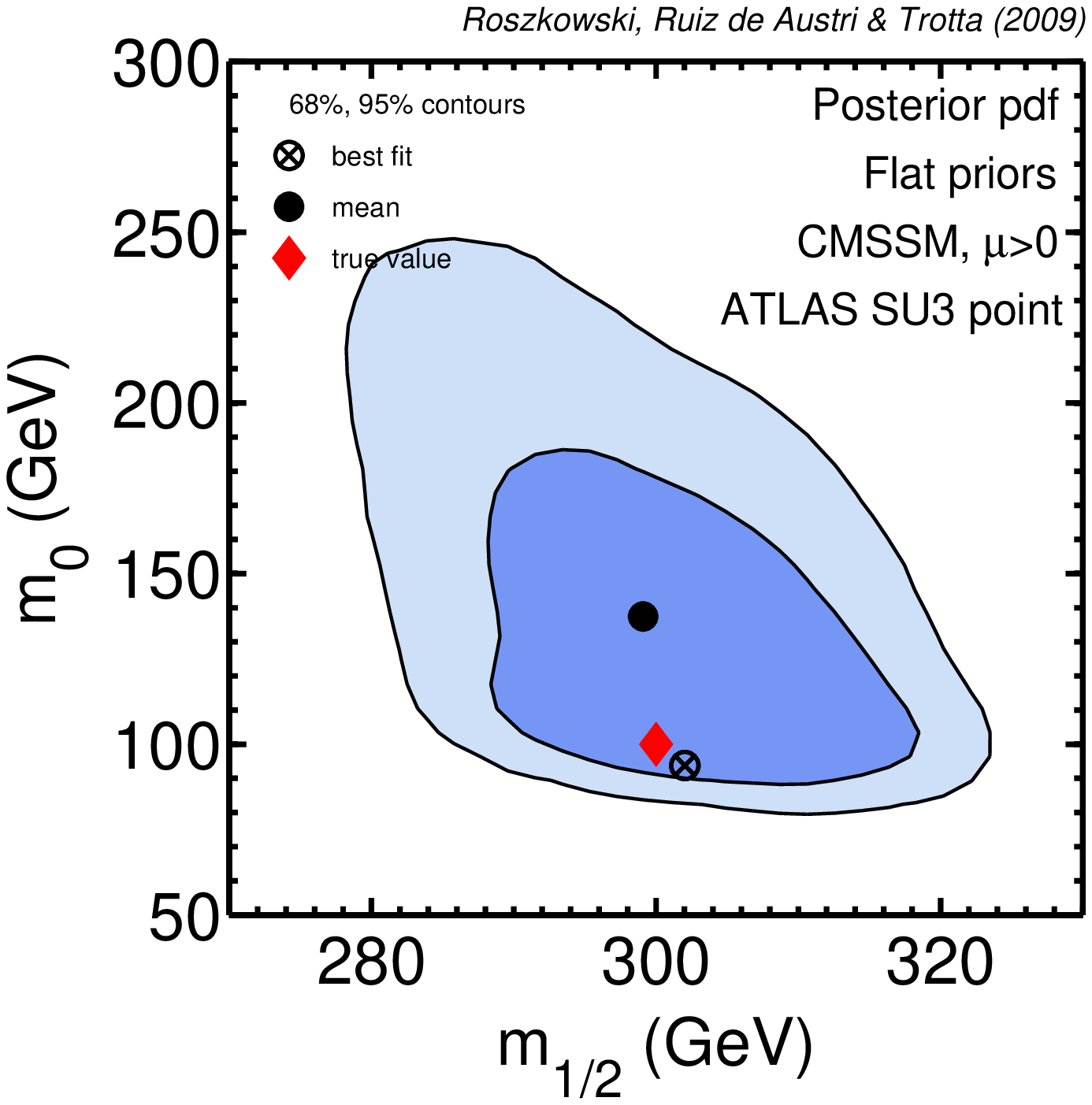} 
\includegraphics[width=\ww]{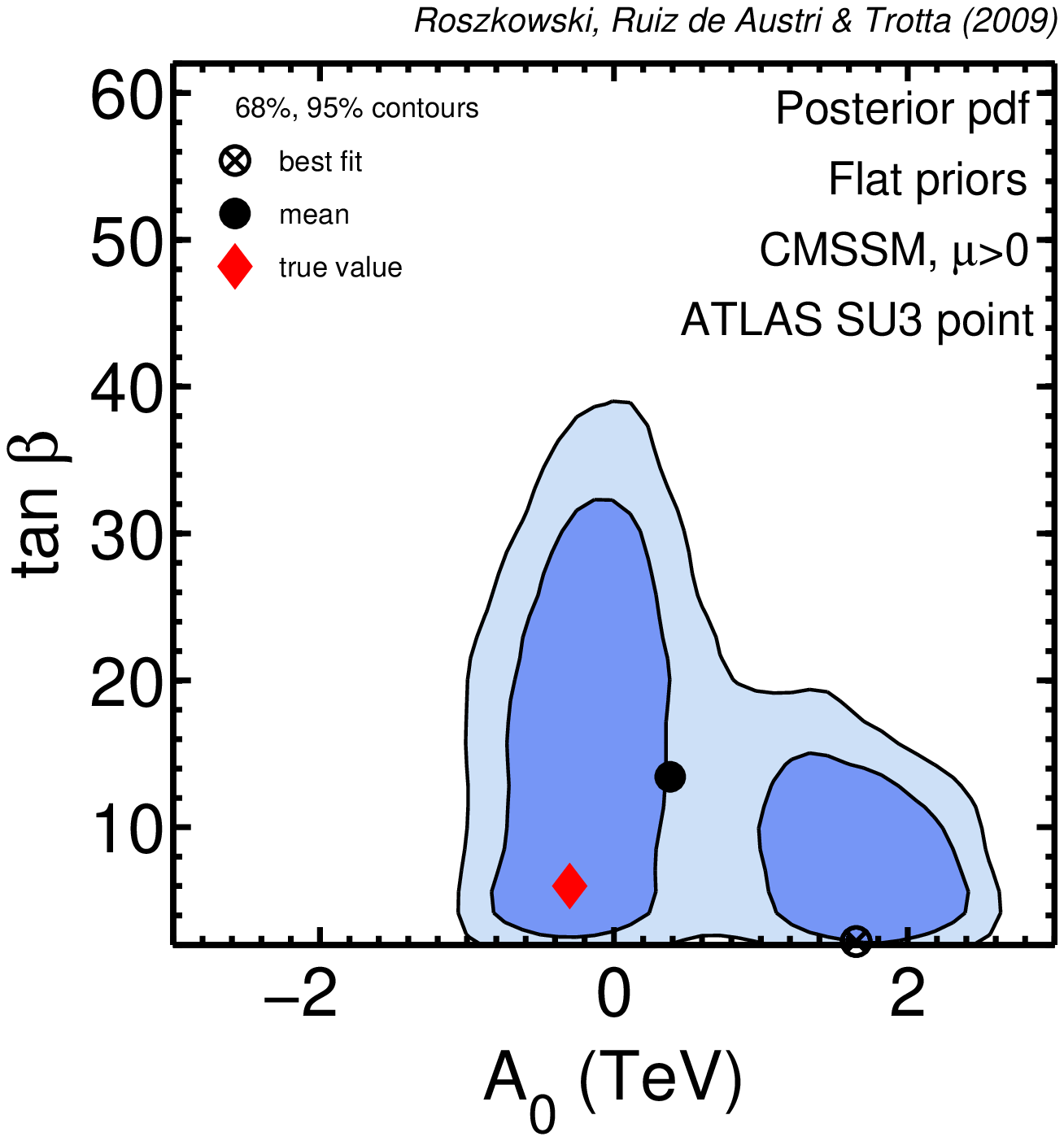} \\
\end{center}
\caption[test]{The same as in Fig.~\protect\ref{fig:CMSSM_LHC_2D_log}
  but for the flat prior case.  Notice that the two choices of priors
  lead to statistically indistinguishable results. }
\label{fig:CMSSM_LHC_2D_flat}
\end{figure}

\subsection{Using ATLAS data only and non-informative priors}\label{sec:atlasonly}

\begin{figure}[tbh!]
\begin{center}
\includegraphics[width=\ww]{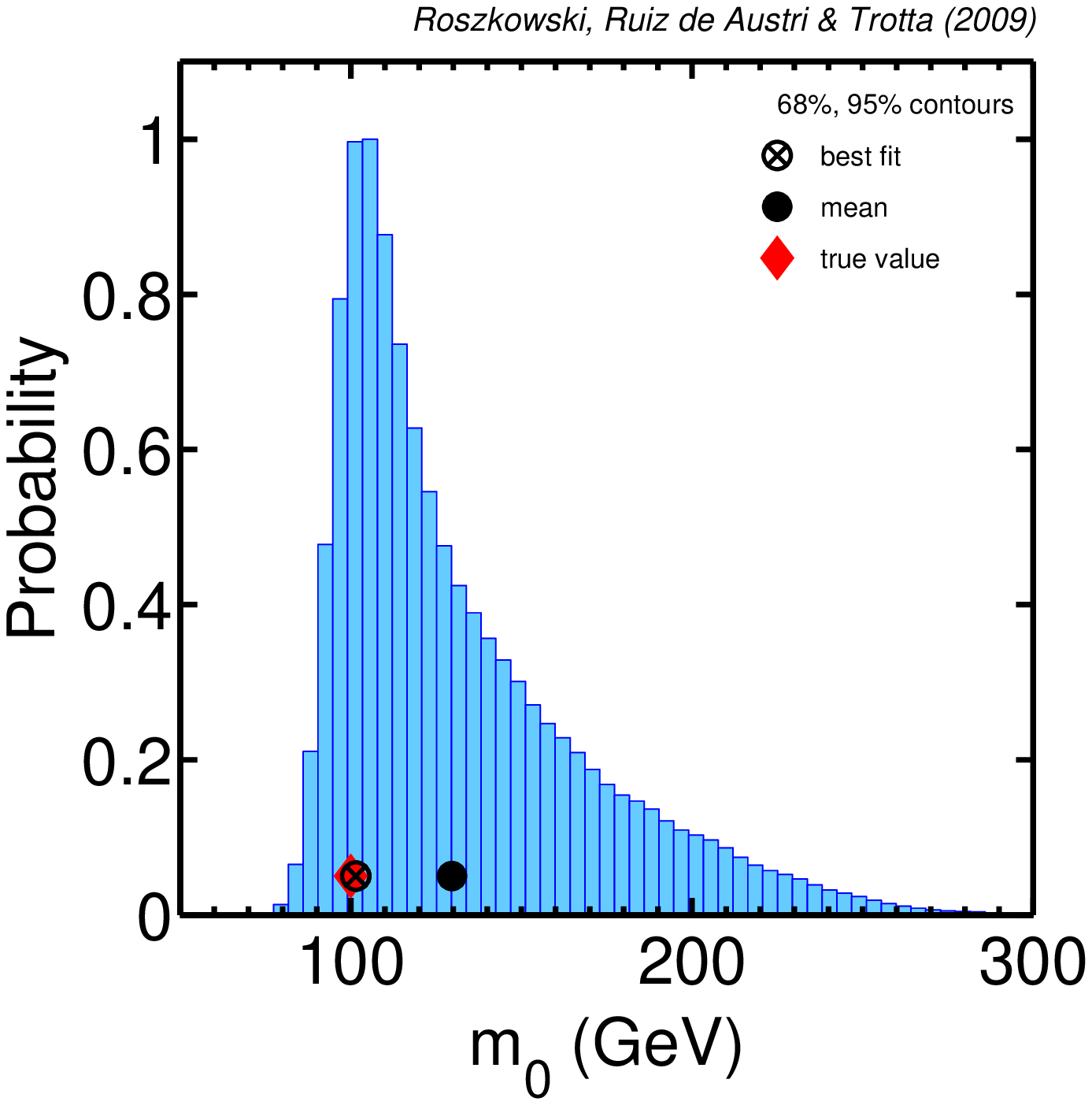}
\includegraphics[width=\ww]{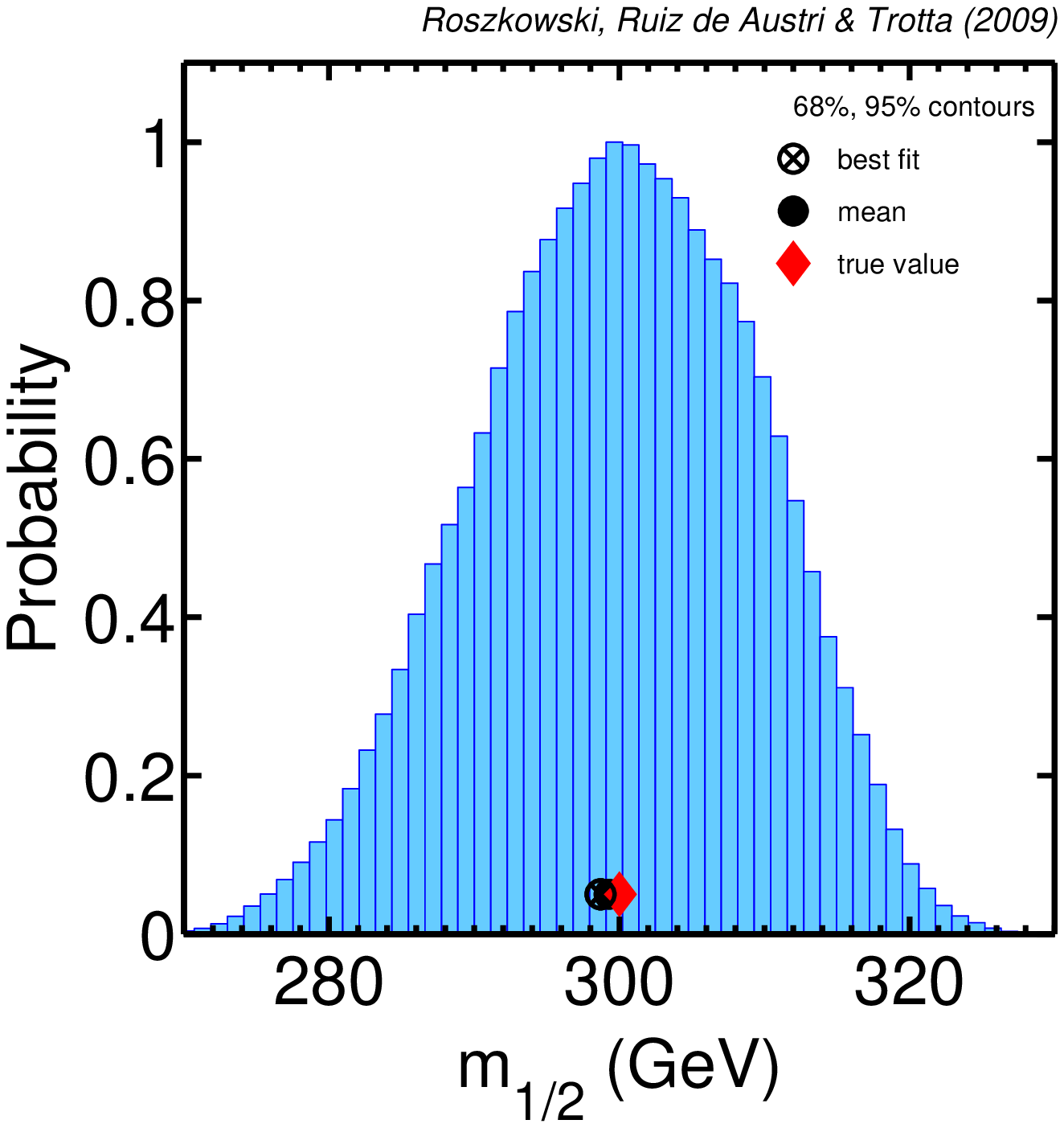} \\
\includegraphics[width=\ww]{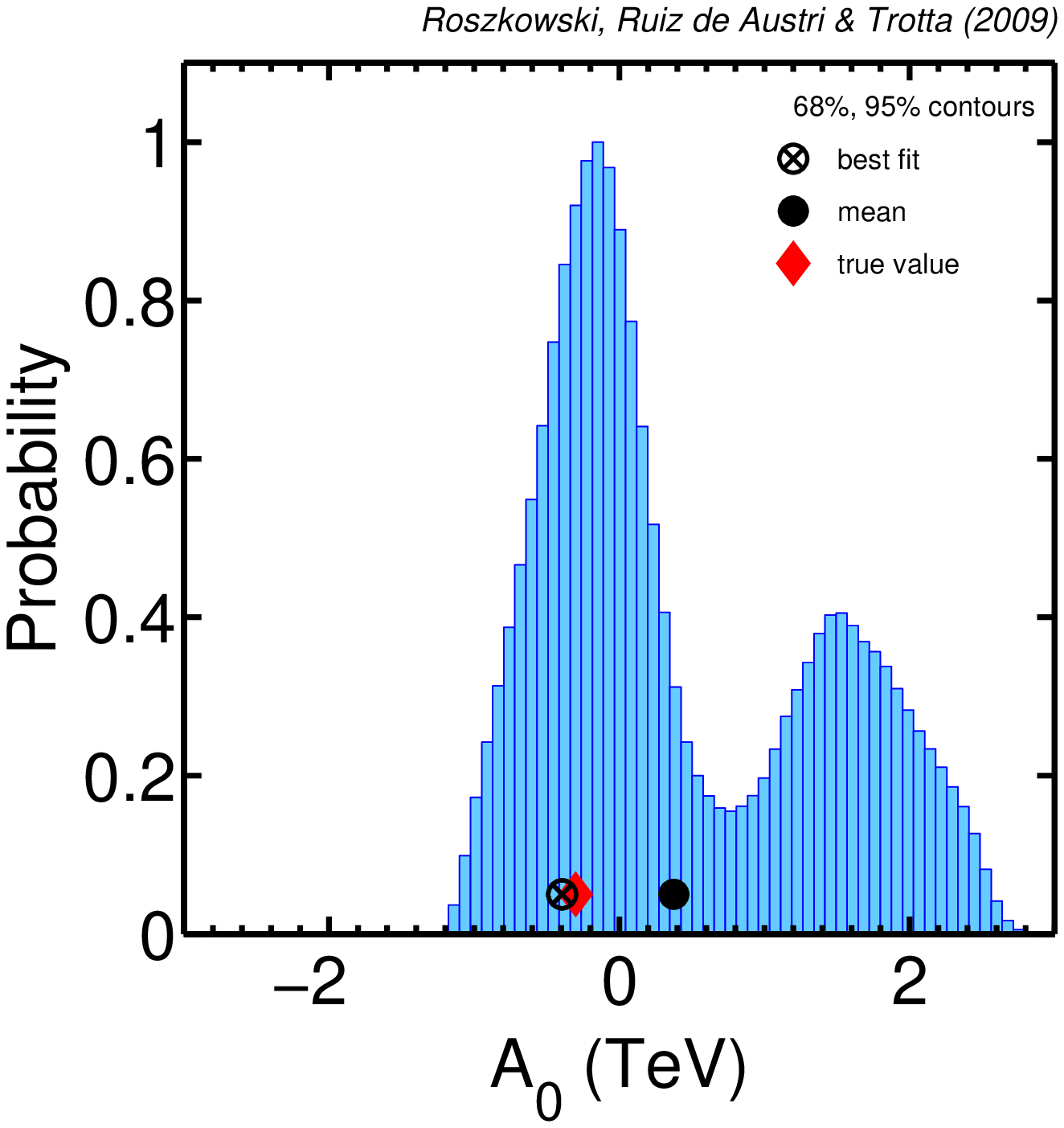}
\includegraphics[width=\ww]{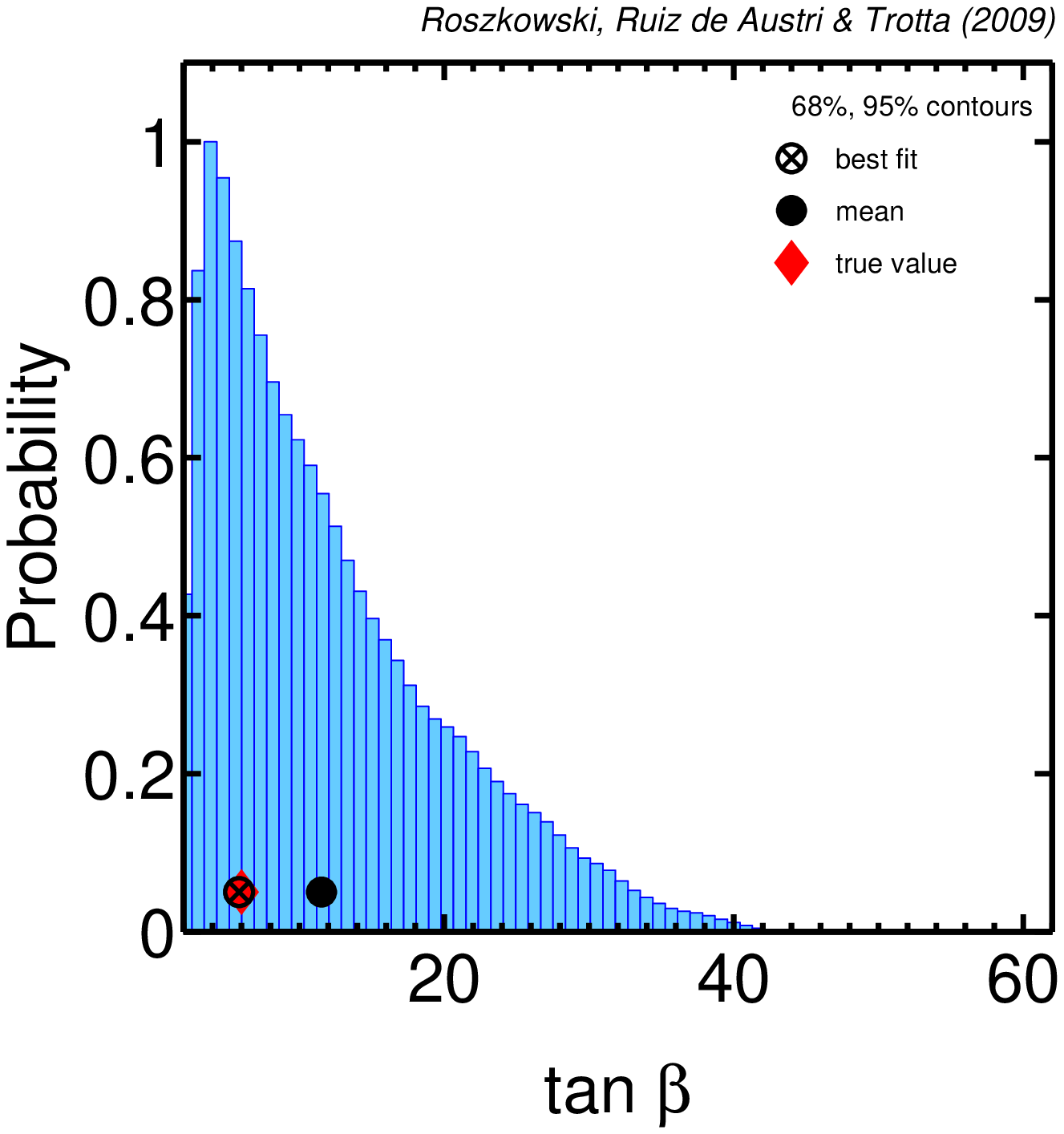}
\caption[test]{1D posteriod pdf for the case of applying only ATLAS mass
  spectrum data, summarized in 1D projections. We show only the
  log prior case, for the flat prior case is essentially
  identical. It is clear that ATLAS data alone is not sufficient
  to reconstruct all of the CMSSM parameters. In particular, while
  $\mhalf$ is well measured, $\azero$ and $\tanb$ remain largely
  undetermined.  }
\label{fig:CMSSM_LHC_1D}
\end{center}
\end{figure}

We begin by considering the constraining power on the CMSSM parameters
of ATLAS mass spectrum data alone. In
\figs{fig:CMSSM_LHC_2D_log}{fig:CMSSM_LHC_2D_flat} we present
2-dimensional (2D) Bayesian posterior pdfs
assuming the log and the flat prior, respectively, while the
corresponding 1D pdfs for the log prior case are shown in
\fig{fig:CMSSM_LHC_1D}.  By examining the $(\mhalf,\mzero)$ plane for
the log prior in \fig{fig:CMSSM_LHC_2D_log}, we can see that both the
68\% (inner regions) and the 95\% (outer regions) total probability
regions are well centered around the true value. Our algorithm
recovers the best-fit point within the limits of numerical noise for
all CMSSM parameters, except for $\azero$, where the reconstructed
best-fit ends up in a wrong region of the parameter space due to a
fundamental degeneracy discussed below. The posterior mean is also
reasonably close to the true value (within 1$\sigma$ for all
parameters, except for $\azero$), although it is slightly skewed due
to the asymmetric nature of the contours, which exhibit heavier tails
than Gaussian (see also Fig.~\ref{fig:CMSSM_LHC_1D} below).  On the
other hand, $\tanb$ is somewhat less well reconstructed, yielding only
an upper limit.

In contrast, $\azero$ is rather poorly constrained in this case, and
actually shows a sign ambiguity.  This is because it enters the
analysis in a rather indirect way, mostly via the off-diagonal terms
$X_\tau=A_\tau -\mu \tanb$ in the stau mass matrix, where $A_\tau$ is
the value of $\azero$ at the EW scale evaluated with its RGE and $\mu$
is computed from the usual requirement of correct electroweak symmetry
breaking.  A closer examination reveals that, for $\azero\sim1\tev$
(in between the two $1\sigma$ regions in the right panel of
Fig.~\ref{fig:CMSSM_LHC_2D_log}), $X_\tau$ is minimized and the mass
difference between $\stautwo$ and $\stauone$ (which plays the r\^{o}le
of the lightest slepton in the decay chain) goes to zero. Since in the
ATLAS analysis only $\stauone$ was considered, such cases are not
allowed. Our study thus reveals that in studying the decay $\chi^0_2
\rightarrow \slepton^{\pm} l^{\mp} \rightarrow\chi^0_1 l^+ l^- $ the
exchange of both $\stautwo$ and $\stauone$ should be considered, as
for some values of $\azero$ their masses, and therefore also relative
contributions, may be comparable.

In the case of the flat prior (\fig{fig:CMSSM_LHC_2D_flat}) the emerging
picture remains essentially identical, thus confirming that the prior
choice becomes less of an issue 
once the constraining power of the data is
sufficiently strong, as expected.  Many of the features seen in
\fig{fig:CMSSM_LHC_2D_log} are displayed more clearly in
\fig{fig:CMSSM_LHC_1D} where the corresponding 1D pdfs are presented
for the log prior case only; the flat prior produces very similar
results and is therefore not shown. 
We give the 68\%
and 95\% intervals of our reconstructed CMSSM parameters in
Table~\ref{tab:reconstruction}. A comparison with the profile
likelihood is carried out further below.

\begin{table}[tbh!]
\centering   
\begin{tabular}{| l | l | l |l| }
\hline
\multicolumn{4}{|c|}{Applying ATLAS data only} \\ \hline
Parameter  &  True value & Best fit & 68\% (95\%) range  \\\hline
$\mhalf$ (GeV) & 300  &  300.4 & \interv{288.2}{308.4}{278.2}{316.3}  \\
$\mzero$ (GeV) & 100 & 104.3  &  \interv{98.7}{173.6}{89.8}{235.2}	 \\ 	
$\tanb$ & 6.0 &  3.8 & $<13.8$ ($<27.4$) (1 tail)  \\
$\azero$ (GeV) & -300  & 1749.7 &\interv{-568.3}{1701.8}{-995.1}{2311.6} \\ \hline
$\mneutone$ (GeV) & 117.9 & 116.9 &\interv{113.7}{120.8}{110.6}{123.7} \\
$\abundchi$  & 0.2332 &  0.2330 &  \interv{0.2264}{0.2870}{0.2096}{0.3450} \\
$\log\sigsip$ (pb) & -8.92 & -8.87 &\interv{-9.14}{-8.42}{-9.45}{-8.04} \\\hline\hline
\multicolumn{4}{|c|}{ Applying ATLAS+WMAP-like data} \\ \hline
Parameter  &  True value & Best fit & 68\% (95\%) range  \\\hline
$\mhalf$ (GeV) & 300  &  302.3 & \interv{293.2}{310.7}{285.5}{317.5}  \\
$\mzero$ (GeV) & 100 & 98.3 &  \interv{95.9}{112.2}{90.9}{151.6} \\ 	
$\tanb$ & 6.0 &  5.5 & $<7.3$ ($<16.3$) (1 tail) \\
$\azero$ (GeV) & -300  & -228.2 &\interv{-498.1}{1437.6}{-887.7}{2199.1}\\ \hline
$\mneutone$ (GeV) & 117.9 & 118.6 &\interv{115.1}{121.3}{112.2}{123.8}  \\
$\abundchi$  & 0.2332 &  0.2333 &  \interv{0.2281}{0.2397}{0.2225}{0.2454} \\
$\log\sigsip$ (pb) & -8.92 & -8.85 &\interv{-9.07}{-8.51}{-9.36}{-8.03}  \\\hline\hline
\multicolumn{4}{|c|}{ Applying ATLAS+Planck-like data} \\ \hline
   $\mhalf$ (GeV) & 300  &  300.5 & \interv{295.7}{311.1}{289.0}
 {317.6}  \\
 $\mzero$ (GeV) & 100 & 99.4 &  \interv{95.3}{106.1}{92.0}{115.6}  \\   
 $\tanb$ & 6.0 &  6.1 & $<4.3$ ($<11.3$) (1 tail) \\
 $\azero$ (GeV) & -300  & -257.4 &\interv{-397.5}{1378.7}{-700.1}
 {2045.5}\\ \hline
 $\mneutone$ (GeV) & 117.9 & 118.0 &\interv{115.9}{121.3}{113.3}
 {123.8}  \\
 $\abundchi$  & 0.2332 &  0.2332 &  \interv{0.2327}{0.2338}{0.2322}
 {0.2345} \\
 $\log\sigsip$ (pb) & -8.92 & -8.88 &\interv{-8.99}{-8.56}{-9.20}
 {-8.31}  \\\hline
\end{tabular} 
 \caption{Reconstructed values and errors for the input CMSSM
  parameters and for some key observables. We also give the best fit
  from our scan. The 68\% and 95\% ranges are computed from the
  posterior pdf as shortest intervals around the mean. For
  definiteness, we have employed the log prior scan but the results
  from the flat prior case are essentially identical. } 
\label{tab:reconstruction}
\end{table}

\begin{figure}[tbh!]
\begin{center}
\includegraphics[width=\ww]{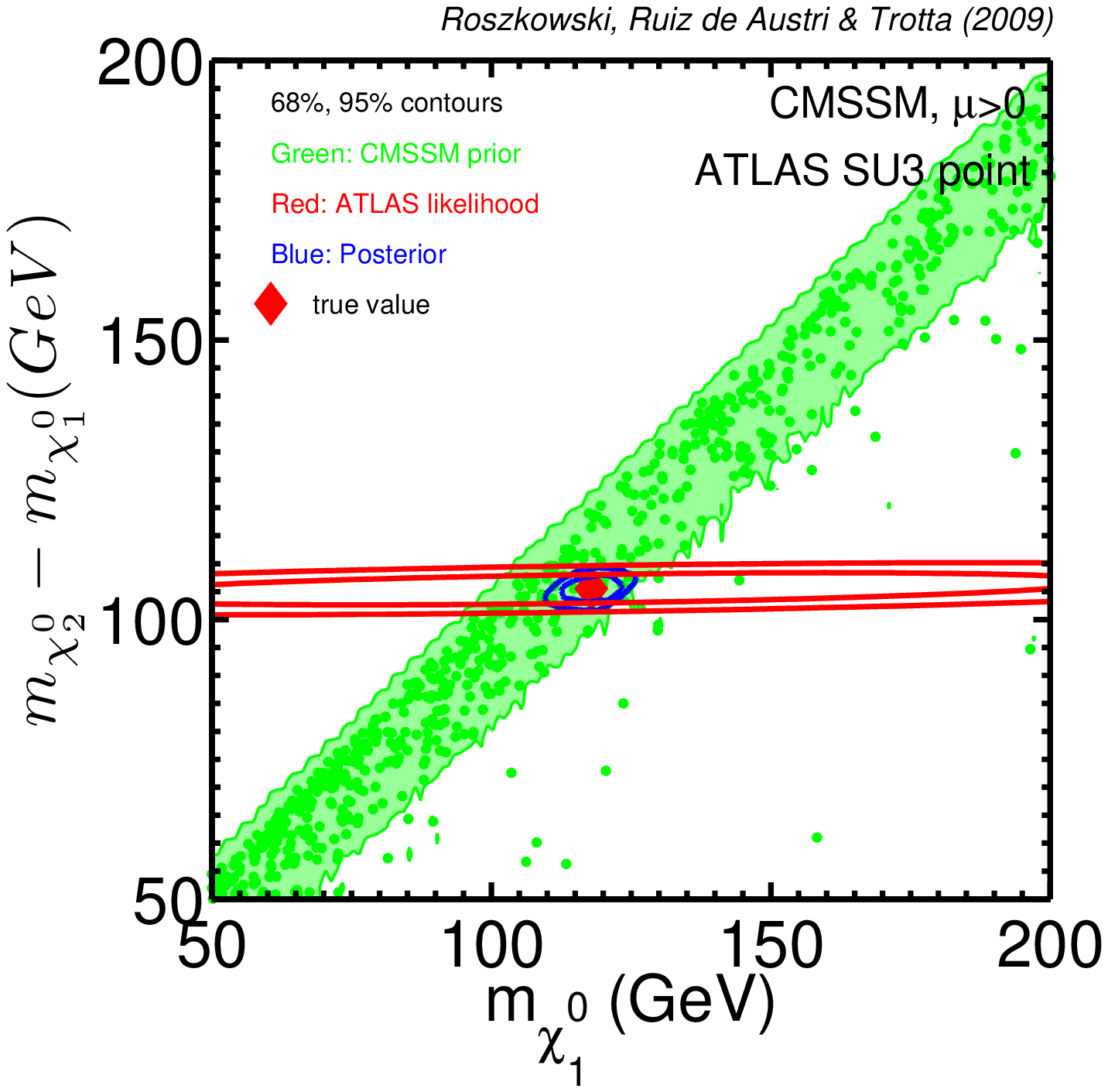}
\includegraphics[width=\ww]{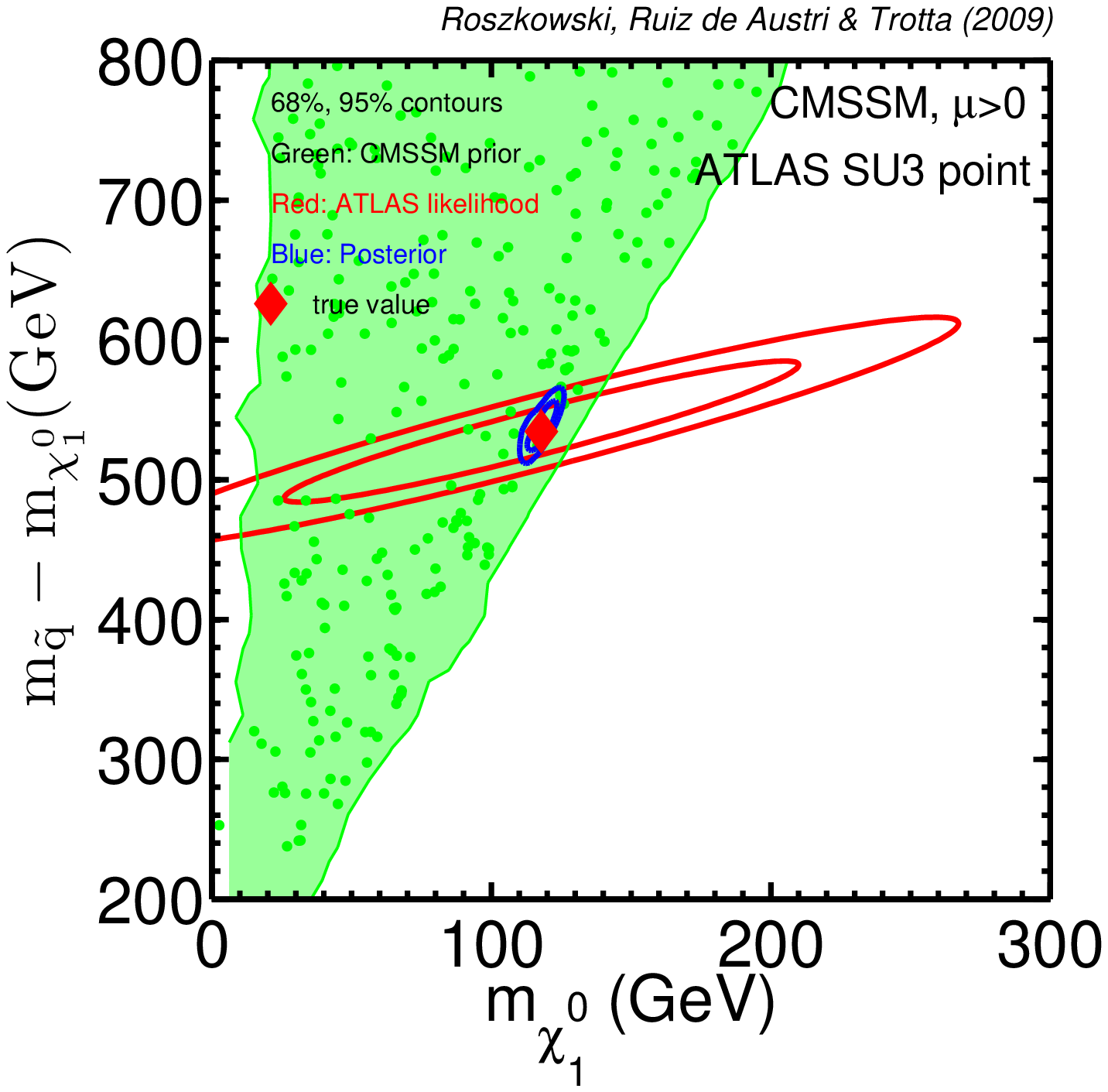}
\caption[test]{Illustration of the extra constraints coming from the
  assumption of the CMSSM as the theoretical framework. The parameter
  space accessible within the CMSSM is given by the green/light gray region (dots
  represents uniformly drawn samples), the red/dark gray (wide) ellipses are the ATLAS
  likelihood (as given by the covariance matrix of
  Table~\ref{tab:covmat}) while the blue/light gray (narrow) contours are the posterior
  constraints. In the context of the CMSSM this allows to derive much
  tighter constraints on $\mneutone$ than it would be possible based
  on the likelihood alone.}
\label{fig:CMSSM_correlation} 
\end{center}
\end{figure}

When considering posterior constraints on the SUSY mass spectrum, it
is apparent that some of the constraints are much stronger than the
likelihood function alone (which actually applies to a more general
case of the MSSM) would seem to imply. For instance, the $1\sigma$
error on $\mneutone$ from the ATLAS likelihood is $60\gev$
(cf.~Table~\ref{tab:covmat}, where the likelihood 1$\sigma$ range is
obtained as the square root of the diagonal elements).  However, the
reconstructed neutralino mass within the CMSSM shows a much smaller
error, of order $\sim 4\gev$, cf.~Table~\ref{tab:reconstruction}.  The
reason for this is that the information supplied by the likelihood is
supplemented by the internal structure of the CMSSM parameter space,
within which the masses of many of the sparticles are highly
correlated. This is demonstrated in Fig.~\ref{fig:CMSSM_correlation},
where one can see that the correlation between masses in the spectrum
within the CMSSM is nearly orthogonal to the constraints provided by
ATLAS for the mass spectrum observables plotted in the Figures. One
can think of this correlation as an additional {\em a priori} piece of
information contained in the model. In other words, given the
theoretical structure of the CMSSM, certain mass combinations in the
spectrum (which are otherwise allowed by the projected ATLAS
constraints, red ellipses in Fig.~\ref{fig:CMSSM_correlation}) are
simply not allowed by the structure of the model. Therefore the final
constraints on the spectrum are much tighter than the likelihood alone
would imply. Supplementing the mass spectrum constraints with a
model-specific implementation, as done here, has the additional
advantage of displaying which part of the constraining power comes
from the experimental data and which one from the theoretical
properties of the model.

\begin{figure}[tbh]
\centering
\includegraphics[width=0.6\textwidth]{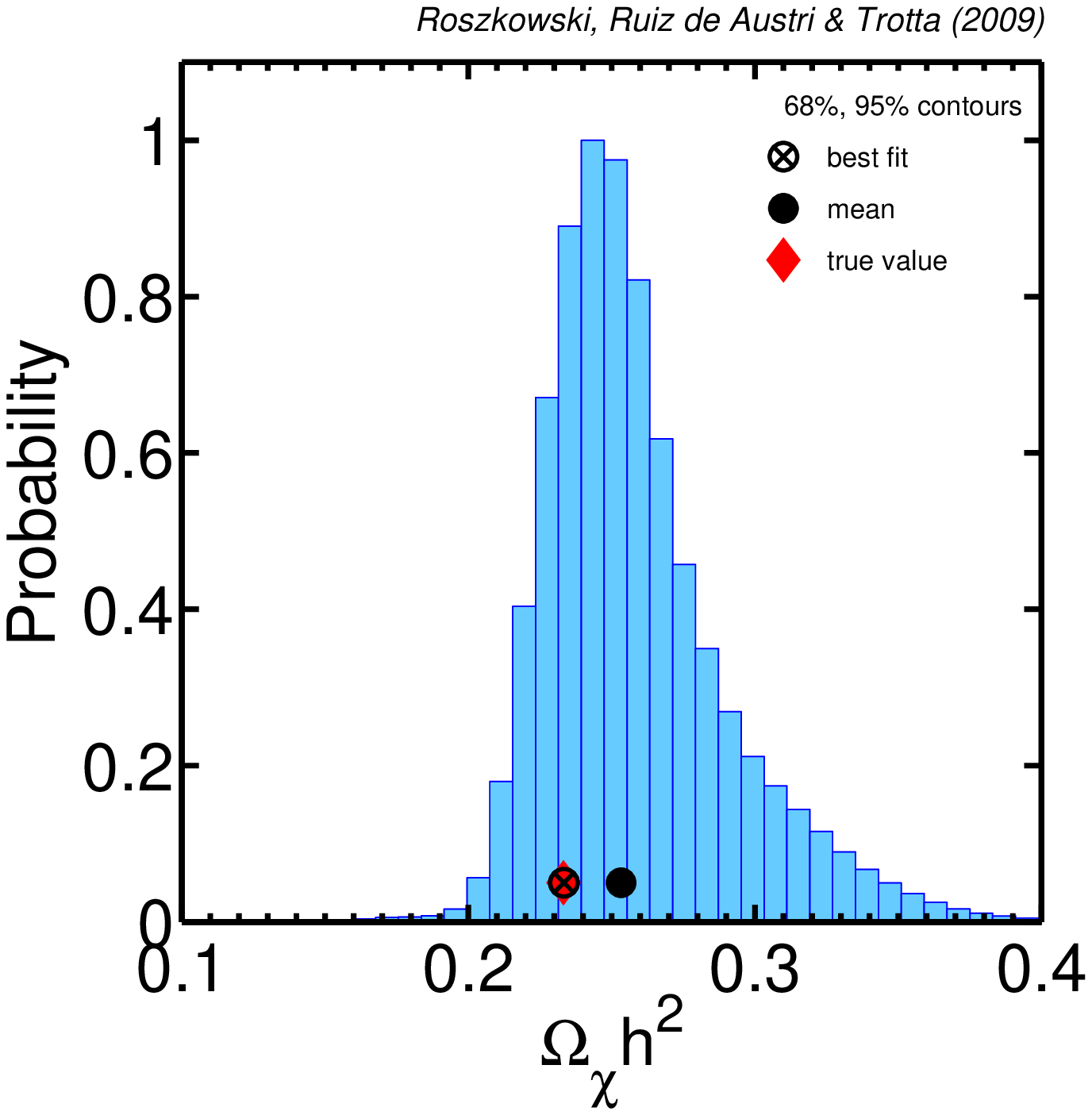}
\caption[test]{1D posteriod pdf for the relic DM abundance $\abundchi$
  of the neutralino, obtained from ATLAS data alone. We show only the
  log prior case, for the flat prior case is basically the same. }
\label{fig:ATLAS_DM} 
\end{figure}

It is interesting to examine how well our procedure allows
one to use the assumed ATLAS data alone to determine
the relic abundance for the ATLAS SU3 point, 
in some analogy with
what, for example, has been done for some other benchmark points in
Refs.~\cite{Nojiri:2005ph,bbpw06}.  This is shown in
\fig{fig:ATLAS_DM} where we find that, from the assumed ATLAS data
alone one would obtain $\abundchi = 0.253 \pm 0.034$, hence with a
relative accuracy of $\sim 13\%$. For this specific point, this would
imply that ATLAS data would determine the neutralino dark matter
abundance at about $7\sigma$. Since the neutralino dark matter
abundance for the ATLAS SU3 benchmark point is some 2.5 times larger
than the value currently preferred by cosmological observations, if we
assume that the accuracy for the benchmark point is representative for
the accuracy that ATLAS will actually find around a point with the
correct cosmological relic abundance of about $0.11$, our estimate is
that ATLAS data alone would be able to determine the DM relic
abundance at the level of $\sim 3\sigma$.
Finally, in our present analysis we have ignored any
theoretical error in the DM abundance prediction. While for the ATLAS SU3
point, which falls into the bulk region, such an error is likely
to be very small, in general it should be folded in when producing the
posterior shown in Fig.~\ref{fig:ATLAS_DM}.  

\subsection{Comparison with the ATLAS Collaboration results}

It is instructive to compare
\figs{fig:CMSSM_LHC_2D_log}{fig:CMSSM_LHC_2D_flat} with Fig.~12 in
Section 9.3 of the ATLAS Report~\cite{atlas09} where 2D marginal
Bayesian posteriors are presented following the ATLAS MCMC
analysis.\footnote{Although it is labeled ``likelihood maps'', the
quantity plotted in Fig.~12 in of Ref.~\cite{atlas09} is actually a
marginal Bayesian posterior (Peter Wienemann - Private Communication),
analogous to the one plotted in our
\figs{fig:CMSSM_LHC_2D_log}{fig:CMSSM_LHC_2D_flat}.} The overall shape
of the high-probability $(\mhalf,\mzero)$ and $(\tanb,\azero)$ regions
is qualitatively similar although quantitatively we find somewhat less
stringent bounds.  In particular, we can see the largest difference in
the case of $\azero$ where the highest probability region found
in~\cite{atlas09} lies on the boundary of the correct region found in
our analysis, while the other, multi-TeV region, is in
Ref.~\cite{atlas09} absent altogether. There is also some difference
in $\mzero$ which in our case is not as well constrained as in
Ref.~\cite{atlas09}.

It is however difficult to carry out a closer comparison, since not
many details are given regarding the setup used in the ATLAS fitting
analysis, in particular, about their treatment of SM nuisance
parameters. Also, the ATLAS fitting analysis was performed directly
from end-point measurements while we used a Gaussian approximation to
the likelihood for masses and mass differencies alone, thus inevitably
loosing a certain amount of information contained in the full
analysis. It is, however, certainly encouraging that our ``shortcut''
method of reconstructing CMSSM parameters using a relatively crude Gaussian approximation to the full ATLAS analysis was able to recover quite
compatible regions of SUSY parameters around their true values. The
only exception is $\azero$, as explained above. As we show below,
adding cosmological relic abundance constraints does help in further
tightening some of the constraints. We conclude that, despite those
differences, overall we find a reasonably good agreement with the
ATLAS analysis. This suggests that not too much information is lost by
carrying out the analysis employing an effective likelihood at mass
spectrum level.

The advantage of our procedure is that it allows one to easily change the
model-specific assumptions: if one replaces the CMSSM by another SUSY
model that one is interested in, the analysis can be carried out
without the need of going through the details of detector performance
and obtaining the ATLAS likelihood numerically via Monte Carlo,
thereby strongly reducing the computational requirements. In fact, 
our analysis requires about 24 hours on eight
3GHz processors, and it is therefore relatively computationally
undemanding.  Furthermore, it would be easy to
adapt our method to employ a more complete likelihood function on the
mass spectrum should this become available as part of the data
products released by the LHC Collaborations. This would allow
theoretical studies of the constraints implied on different SUSY
models without the need to reproduce the full detector-specific signal
reconstruction.

\subsection{Impact of including the DM relic abundance}

\begin{figure}[tbh!]
\begin{center}
\includegraphics[width=\ww]{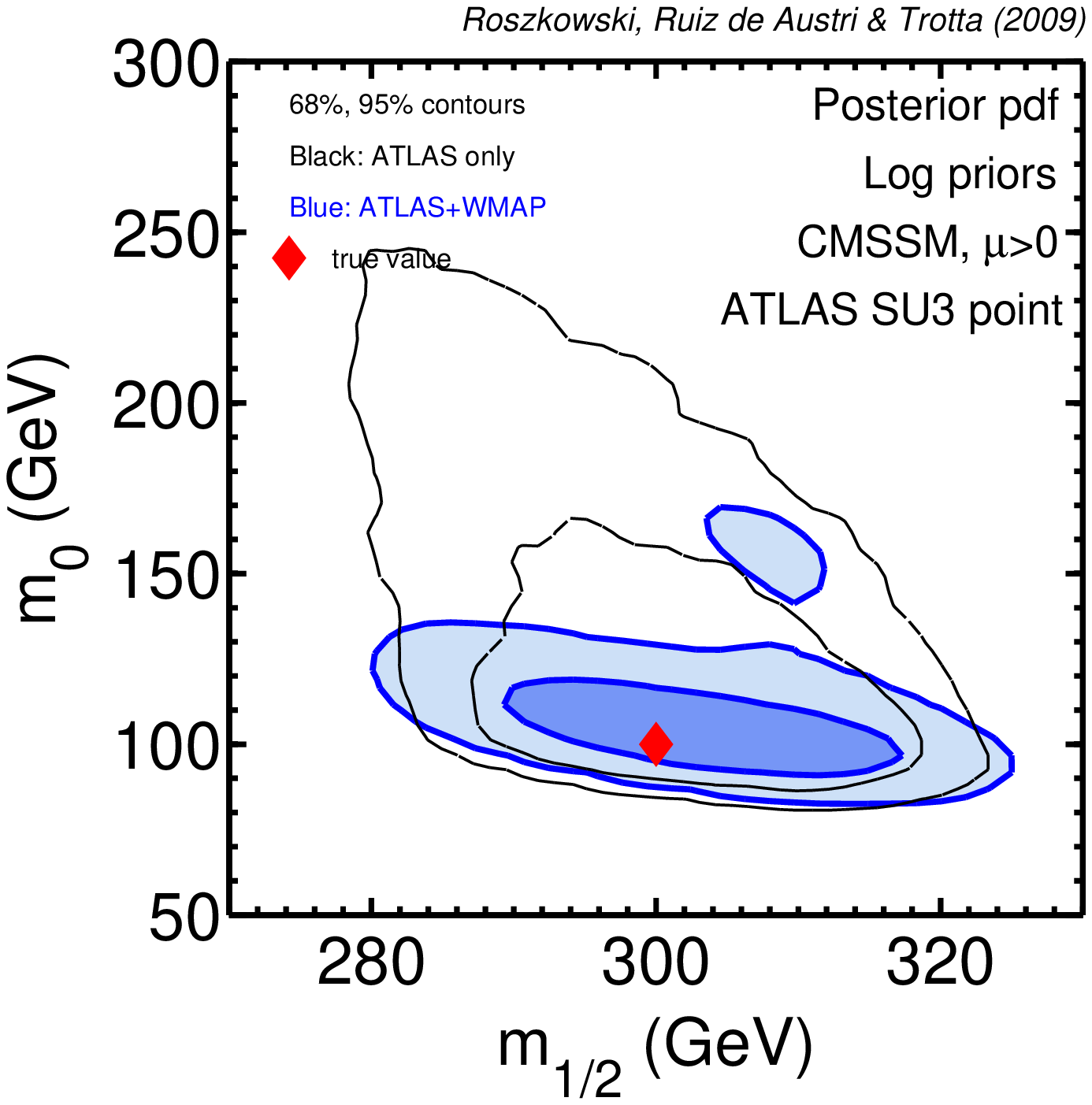}
\includegraphics[width=\ww]{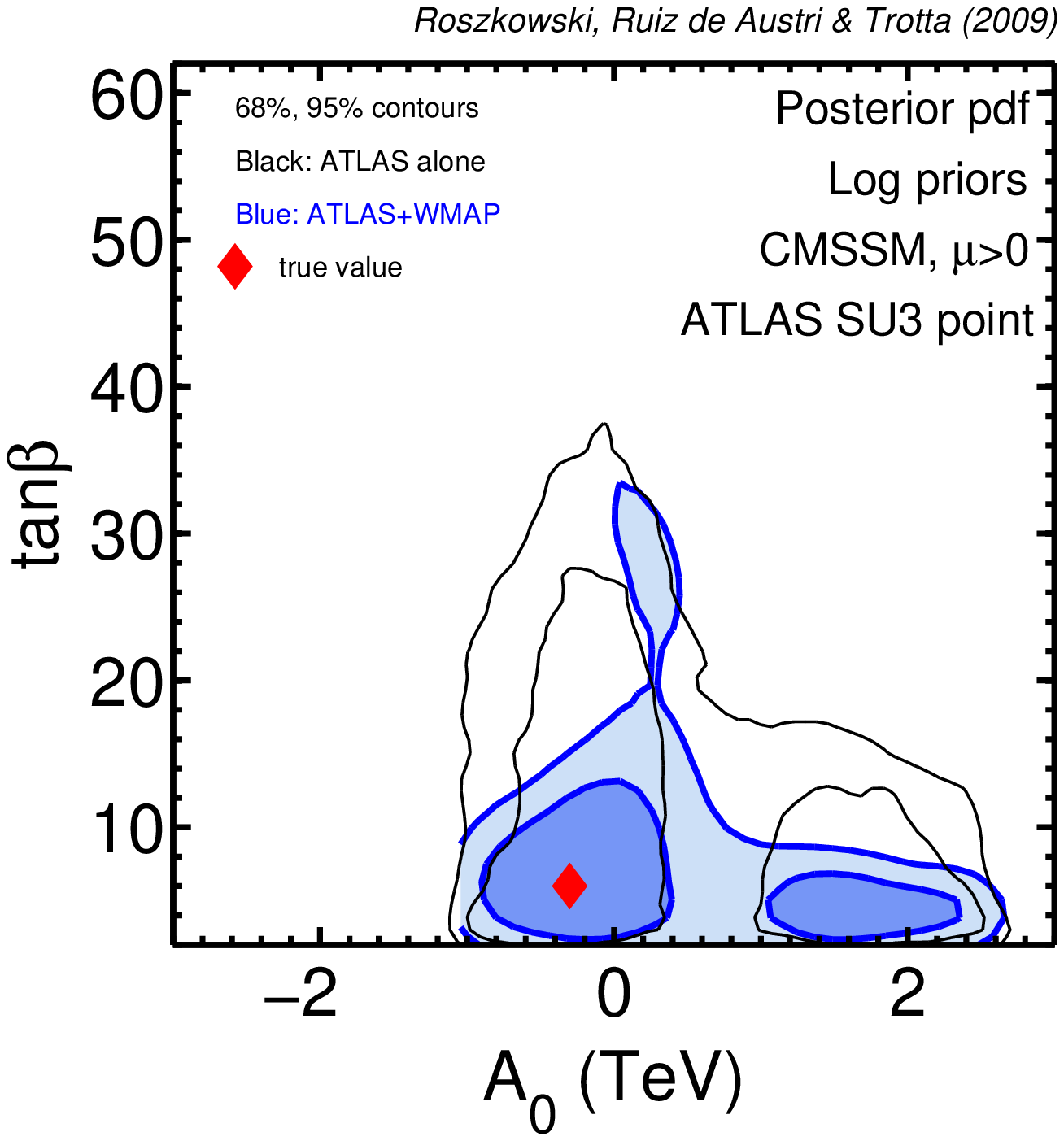}\\
\includegraphics[width=\ww]{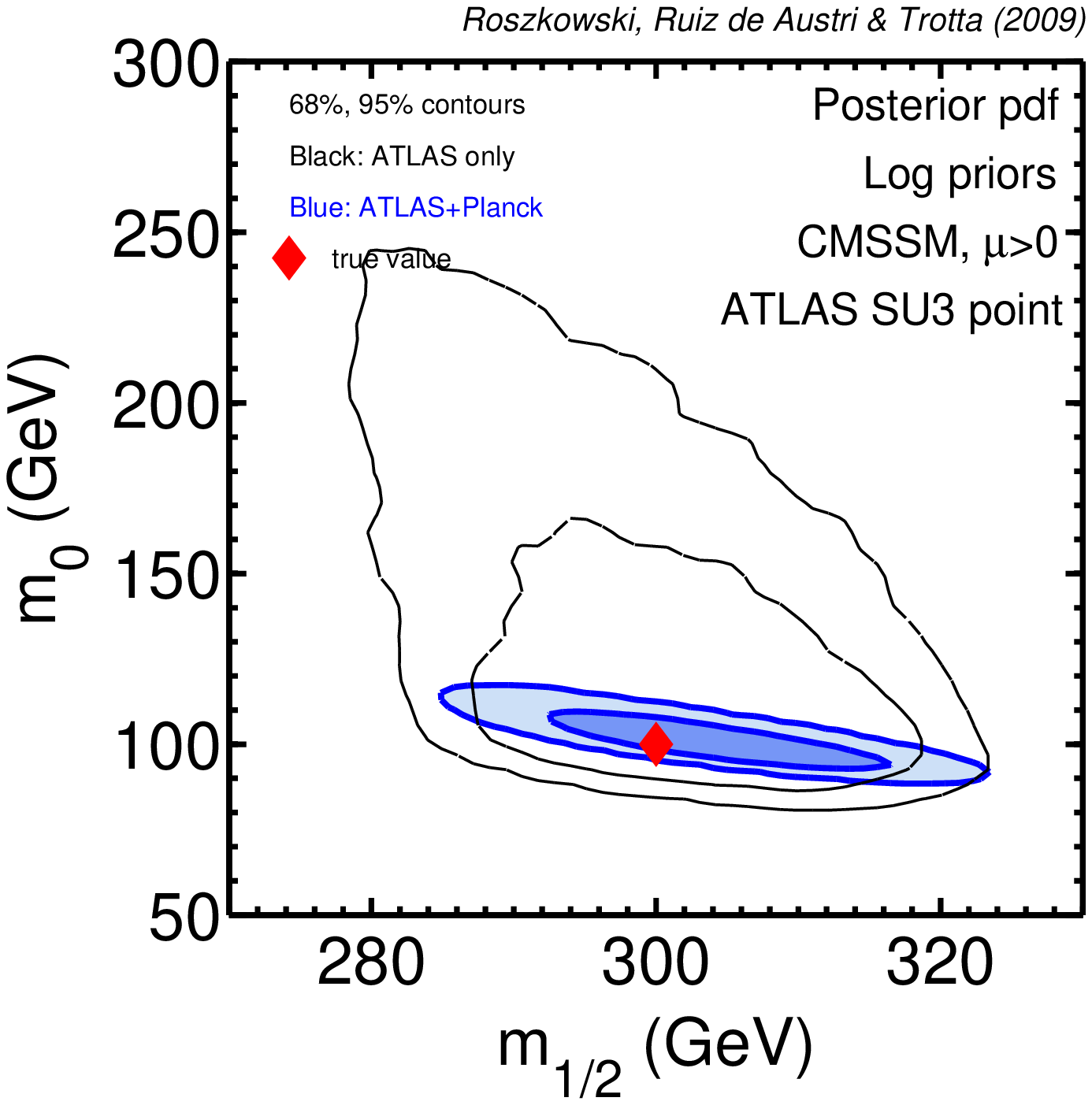}
\includegraphics[width=\ww]{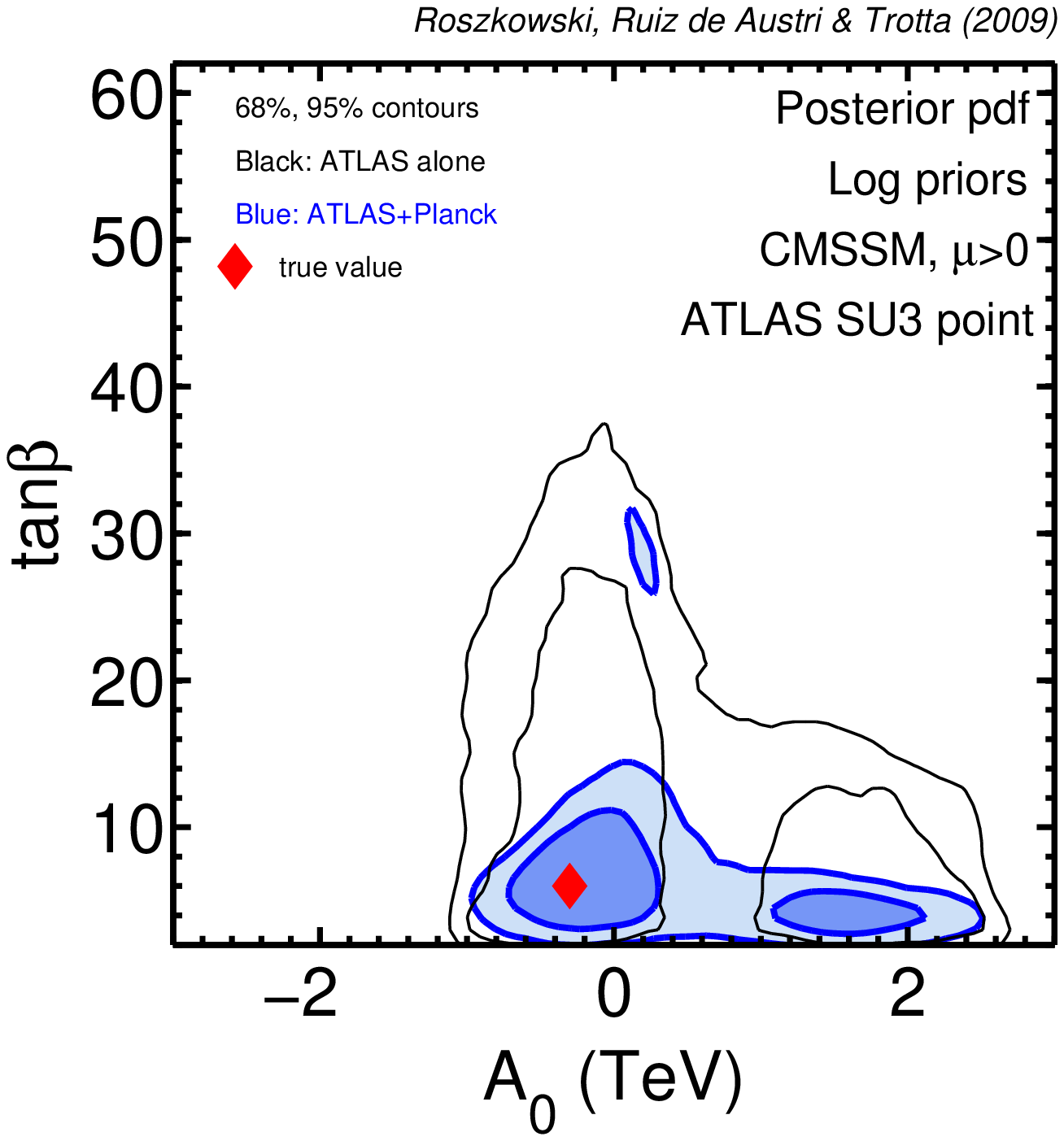}\\
\caption[test]{Impact of adding to the ATLAS data cosmological dark
  matter abundance determination with WMAP-like (upper row) and
  Planck-like (lower row) errors on $(\mhalf,\mzero)$ (left panels)
  and $(\tanb,\azero)$ (right panels). Filled regions are for ATLAS
  plus either WMAP or Planck, while empty contours are for ATLAS
  only. Only the log prior case is presented; the flat one produces
  very similar results. }
\label{fig:DM_impact} 
\end{center}
\end{figure}

We now add to our likelihood function a constraint on the relic
abundance uncertainties, as discussed in Sec.~\ref{sec:likefn}. In
\fig{fig:DM_impact} we show the effect of imposing the ATLAS and WMAP
data (ATLAS+WMAP) in the upper row, and an analogous case for the
ATLAS+Planck case in the lower row. We plot the posterior for the log
prior case; the flat prior case is basically identical. It is clear
that, adding WMAP-like constraints improves the reconstructing power
in determining the CMSSM parameters very considerably in the case of
$\mzero$ (and to some extent also $\tanb$), while the impact on the
other two CMSSM parameters is fairly limited. This can be traced back
to the fact that, in the bulk region, $\abundchi$ is determined
primarily by the mass of the lightest slepton, via a $t$-channel
exchange. Tightening the allowed range of $\abundchi$ selects a more
peaked range of $\mslepton$ and thus also $\mzero$ on which it mostly
depends. On the other hand, $\mhalf$, which primarily determines
$\mchi$, can be adequately constrained already by using only
ATLAS data. As regards $\azero$, the bi-modality still remains as it
is caused by the internal structure of the CMSSM.  On the other hand,
a further improvement of the error on $\abundchi$ to the level
expected from Planck does not seem to improve the situation much
beyond the ATLAS+WMAP case.  In this context we again emphasize that,
at this level of accuracy, it will be essential to achieve a similar,
or better, level of theoretical errors, which may be challenging even
for the bulk region.

\begin{figure}[tbh!]
\begin{center}
\includegraphics[width=\ww]{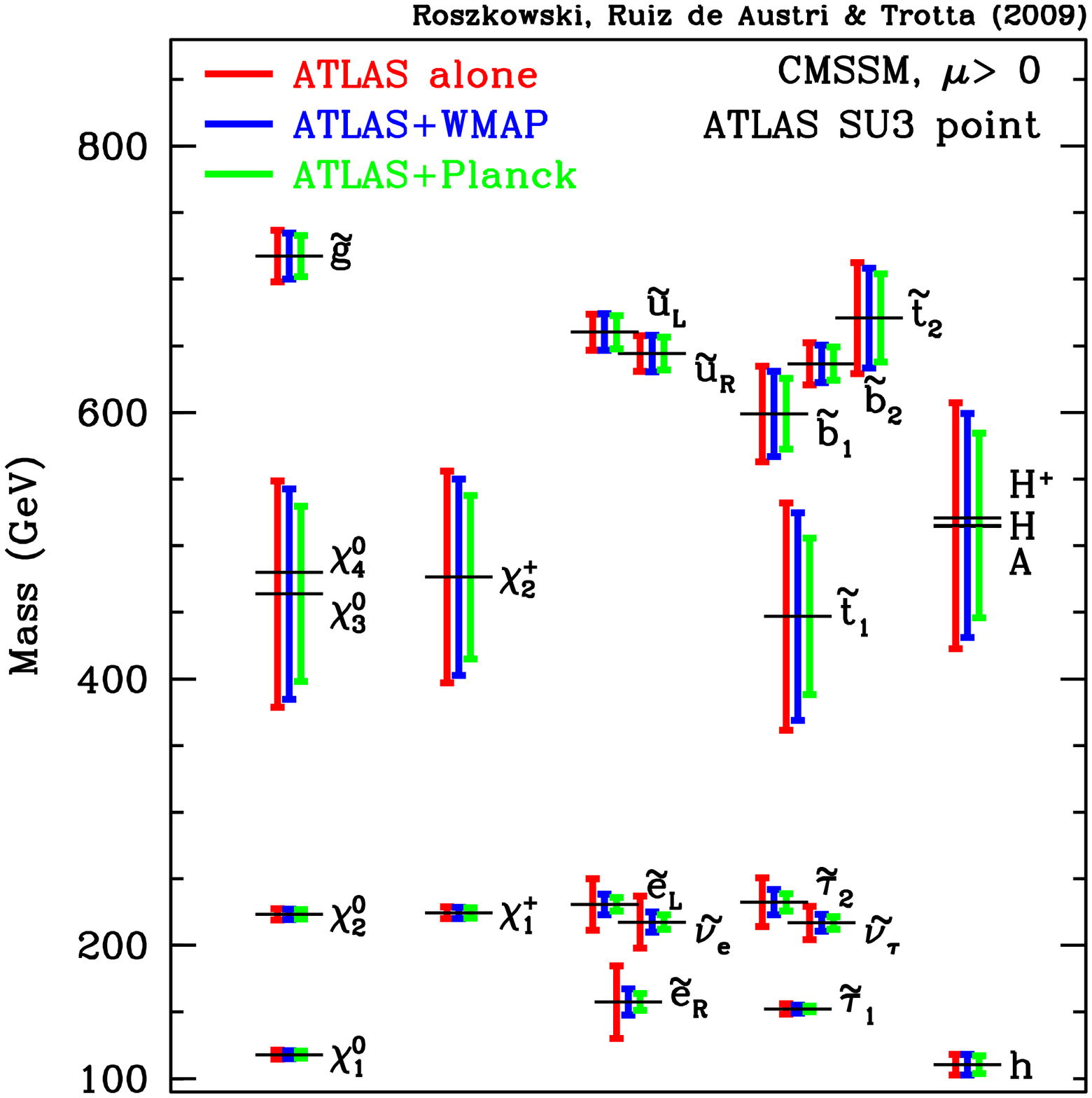}
\caption[test]{Reconstruction of the SUSY mass spectrum using
  projected ATLAS data only (red/leftmost errobar), adding WMAP-like
  constraints on the dark matter relic abundance (blue/central
  errorbar) and adding Planck-like dark matter constraints
  (green/rightmost errorbar). The errorbars represent the 68\% range
  of the Bayesian posterior for the log prior.  (The results for the flat
  prior is essentially identical.)  }
\label{fig:CMSSM_spectrum}
\end{center}
\end{figure}

In \fig{fig:CMSSM_spectrum} we show the constraints on the masses of several
superpartners obtainable with the three sets of
data considered in this paper. We can see that in the case of the gauginos
($\chi^0_{1,2}$, $\charone$ and $\gluino$), whose masses are
determined primarily by $\mhalf$ (which is well reconstructed), the
errors are rather small, while for higgsino-like states
($\chi^0_{3,4}$ and $\chi^\pm_2$) the errors are large because of a poor
determination of the $\mu$ parameter. For the states whose mass
strongly depends on $\mzero$ (spin-zero superpartners) the errors
again reflect that of the common scalar mass, whose reconstruction,
while reasonable, is not as good as for $\mhalf$. 

The impact of further imposing other often used constraints from
$\bsgamma$ and $\gmtwo$ is in the present case rather limited. This is
because the total error in the first quantity is still substantial while the dominant SUSY
contribution to $\gmtwo$ comes from sneutrino-chargino exchange. In 
the low mass region the masses of both particles are low and thus
their contribution can be large enough to significantly reduce the 
discrepancy between the experimental data and the SM value.

\begin{figure}[tbh!]
\begin{center}
\includegraphics[width=\ww]{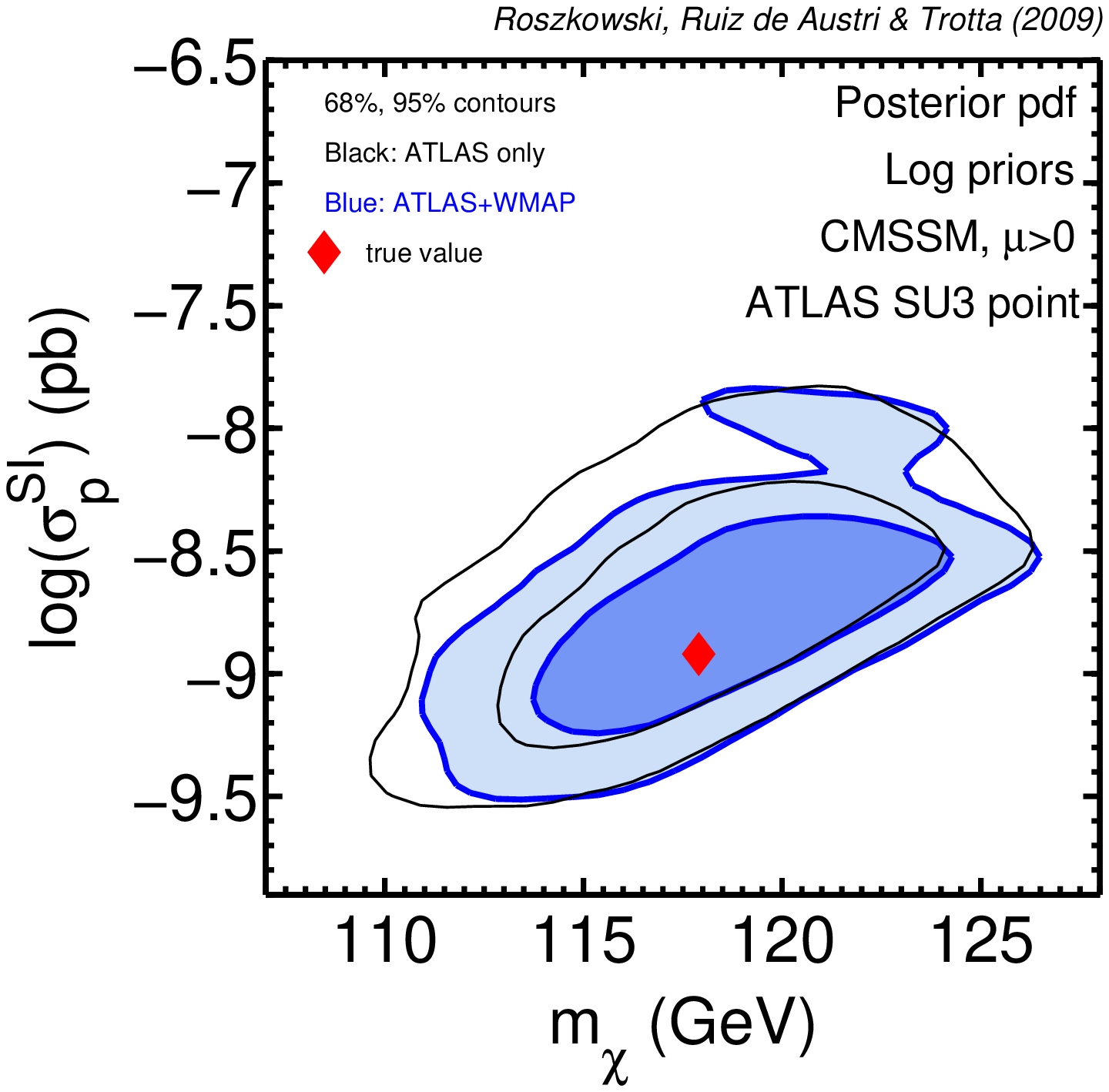}
\includegraphics[width=\ww]{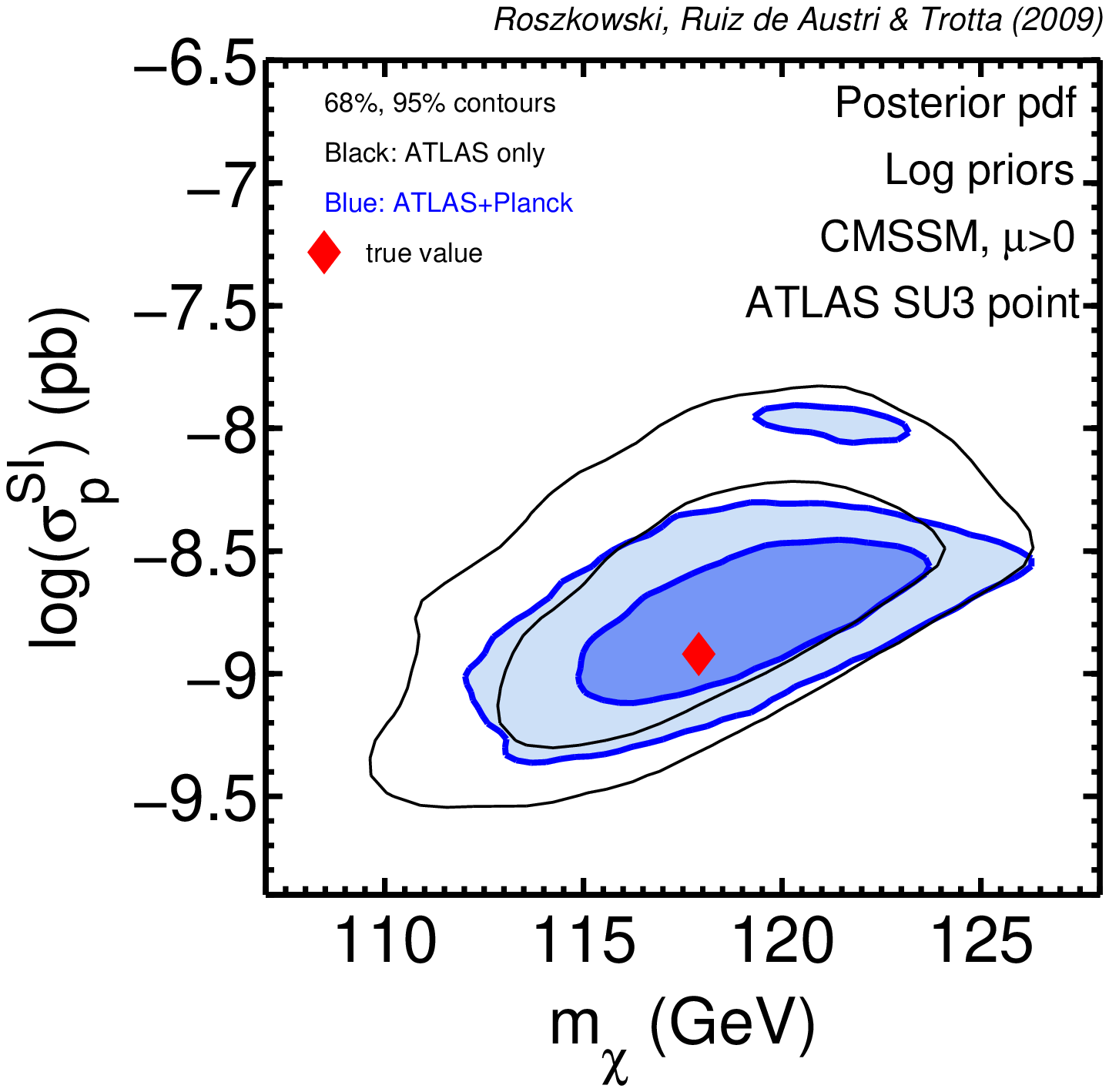}\\
\caption[test]{Impact of adding to the ATLAS data cosmological dark
  matter abundance determination with WMAP-like (left panel) and
  Planck-like (right panel) errors on the predictions for the spin-independent cross
  section of dark matter neutralino scattering off a proton, relevant
  for direct detection 
  experiments.  Filled regions are for ATLAS plus either WMAP or
  Planck, while empty contours are for ATLAS only.}  
\label{fig:DM_impact_DD}
\end{center}
\end{figure}

Finally, we investigate how well one can predict the spin-independent
cross section $\sigsip$ of dark matter neutralino scattering off a
proton tested in direct detection experiments. As can be seen from
\fig{fig:DM_impact_DD}, at 68\% the value of $\sigsip$ will
remain uncertain to within about one order of magnitude, while the
neutralino mass will be very well constrained by LHC data as a
reflection of the bounds on $\mhalf$. This is because in the case
studied here
$\sigsip$ is too a large extent determined by a $t$-channel heavy
scalar Higgs exchange, where, in addition to $\mchi$, the main two
parameters are $\tanb$ and the Higgs mass which shows a considerable
spread of values, mostly due to the larger uncertainty in
$\mzero$. Adding information about the DM relic abundance
therefore improves the situation only in a fairly limited way.

\subsection{Impact of a naturalness prior} \label{sec:naturalness}

We now turn to investigating the impact that a highly informative
prior choice based on naturalness considerations would have on the
Bayesian posterior. The CCR effective prior
implements Occam's razor penalization of regions of the parameter
space exhibiting large fine-tunings~\cite{ccr08}. This implies that the
statistical weight of regions with large $\tanb$ is reduced, since
fine-tuning generally increases with increasing $\tanb$. The same
applies to the soft-terms, except for $\mzero$ where lower fine-tuning
is actually achieved in the TeV range, in the so-called hyperbolic
branch/focus point (FP) region~\cite{Chan:1997bi,focuspoint-fmm}.  

\begin{figure}[tbh!]
\begin{center}
\includegraphics[width=\ww]{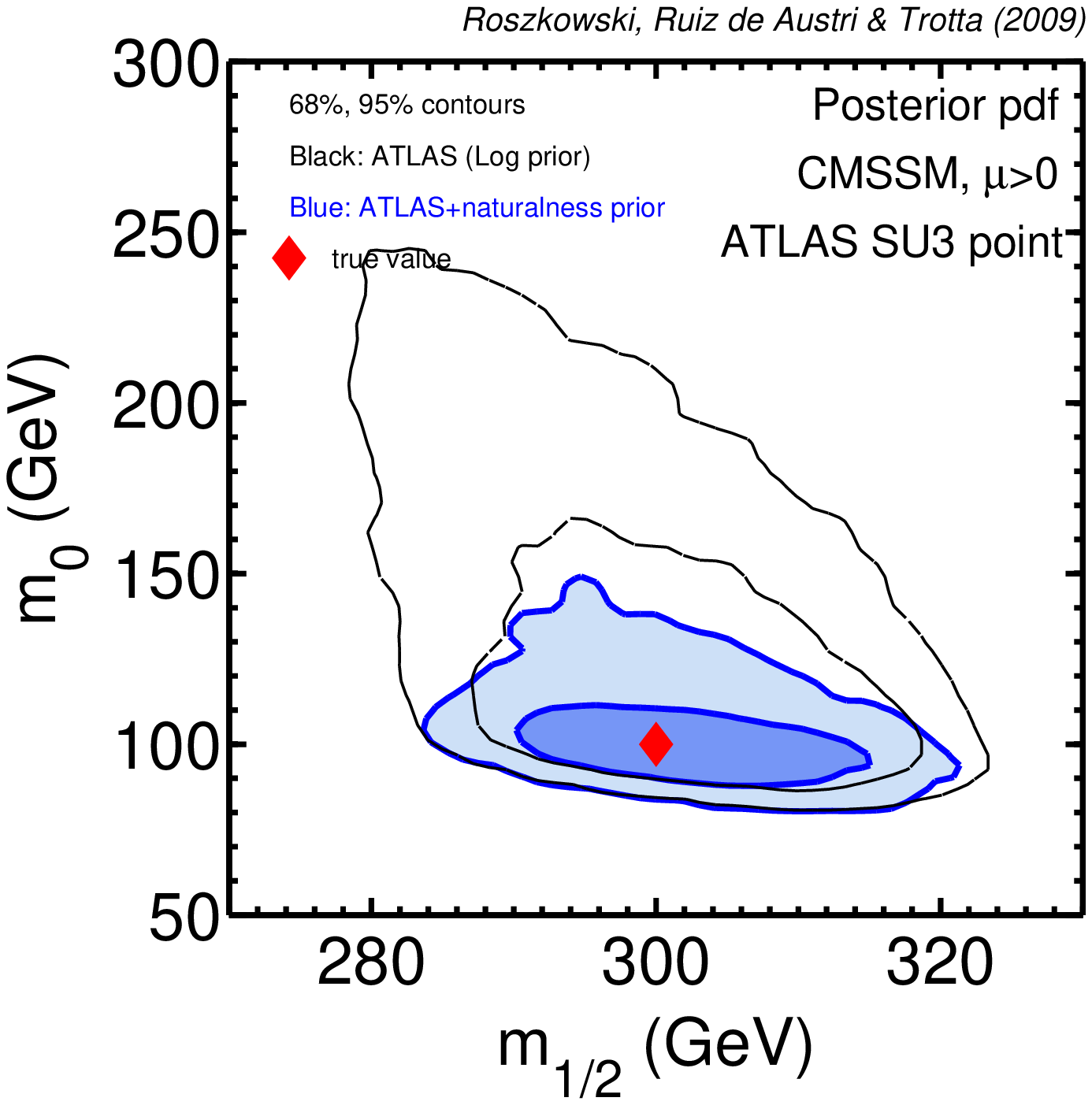}
\includegraphics[width=\ww]{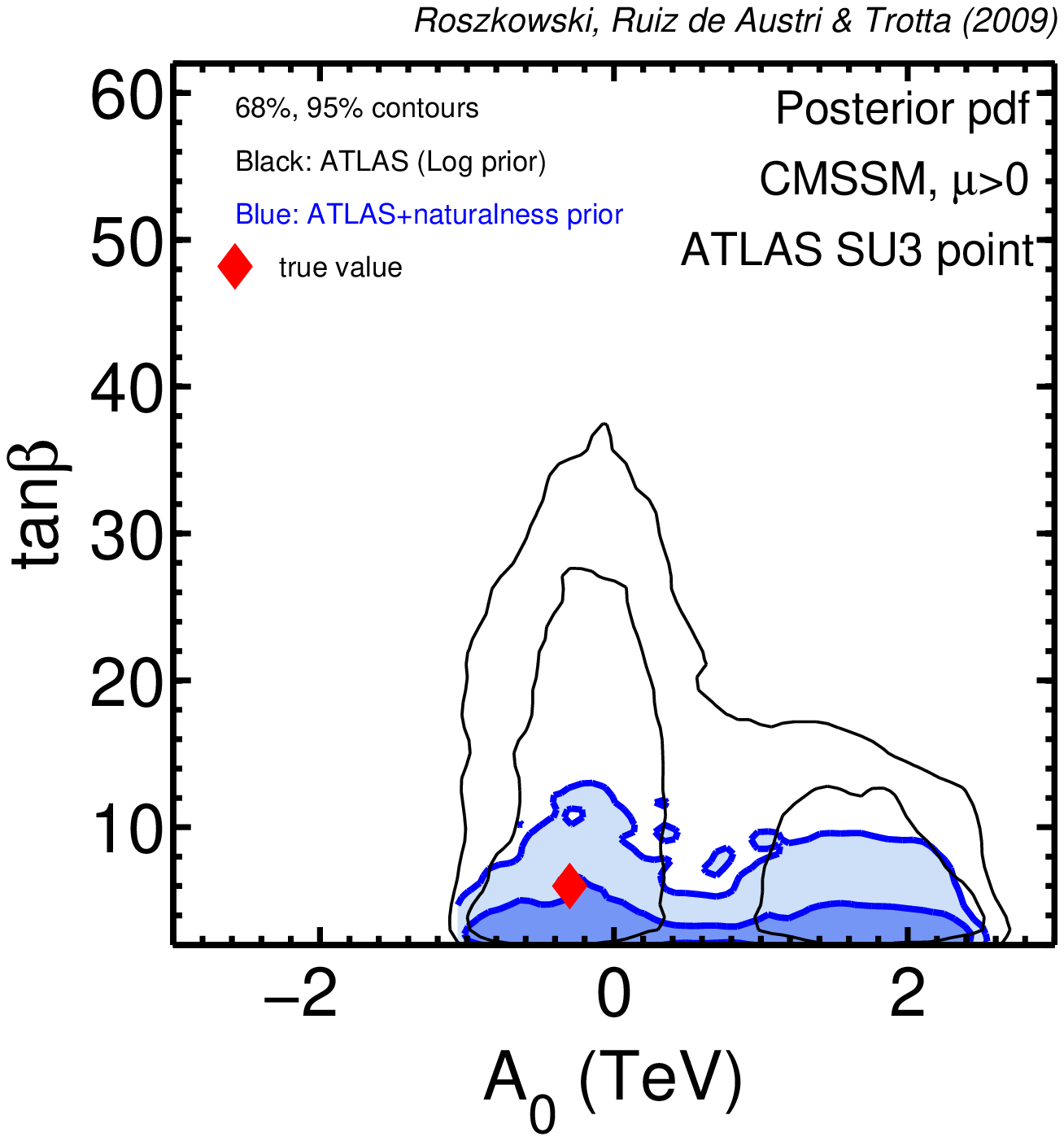} 
\caption[test]{2D posterior pdf for the case of applying ATLAS mass
  spectrum data alone with an informative naturalness prior (the CCR
  prior, filled contours), compared with the posterior obtained using
  non-informative log priors (empty contours). }
\label{fig:CMSSM_CCR_2D}
\end{center}
\end{figure}

The posterior pdf for the CCR prior for the CMSSM parameters is shown
in Fig.~\ref{fig:CMSSM_CCR_2D} as blue-shaded 68\% (darker) and 95\%
(lighter) regions, where for comparison we also show the 
non-informative log prior case (the corresponding black contours).  We
observe that the CCR prior leads to much tighter errors on especially
$\tanb$, and to some extent  also $\mzero$, by assigning a larger penalty, and therefore
stronger constraints, to ``less natural'' ranges of those
parameters. The posteriors for
$\mhalf$ and $\azero$, on the other hand, are only midly affected by
the CCR prior. This is an example of how supplementing the information from the
likelihood with a naturalness prior coming from theoretical prejudice
leads to a posterior which can be significantly different from one
obtained using non-informative log (or flat) prior.

\section{Comparison with profile likelihood}\label{sec:pl}

\begin{figure}[tbh!]
\begin{center}
\includegraphics[width=\ww]{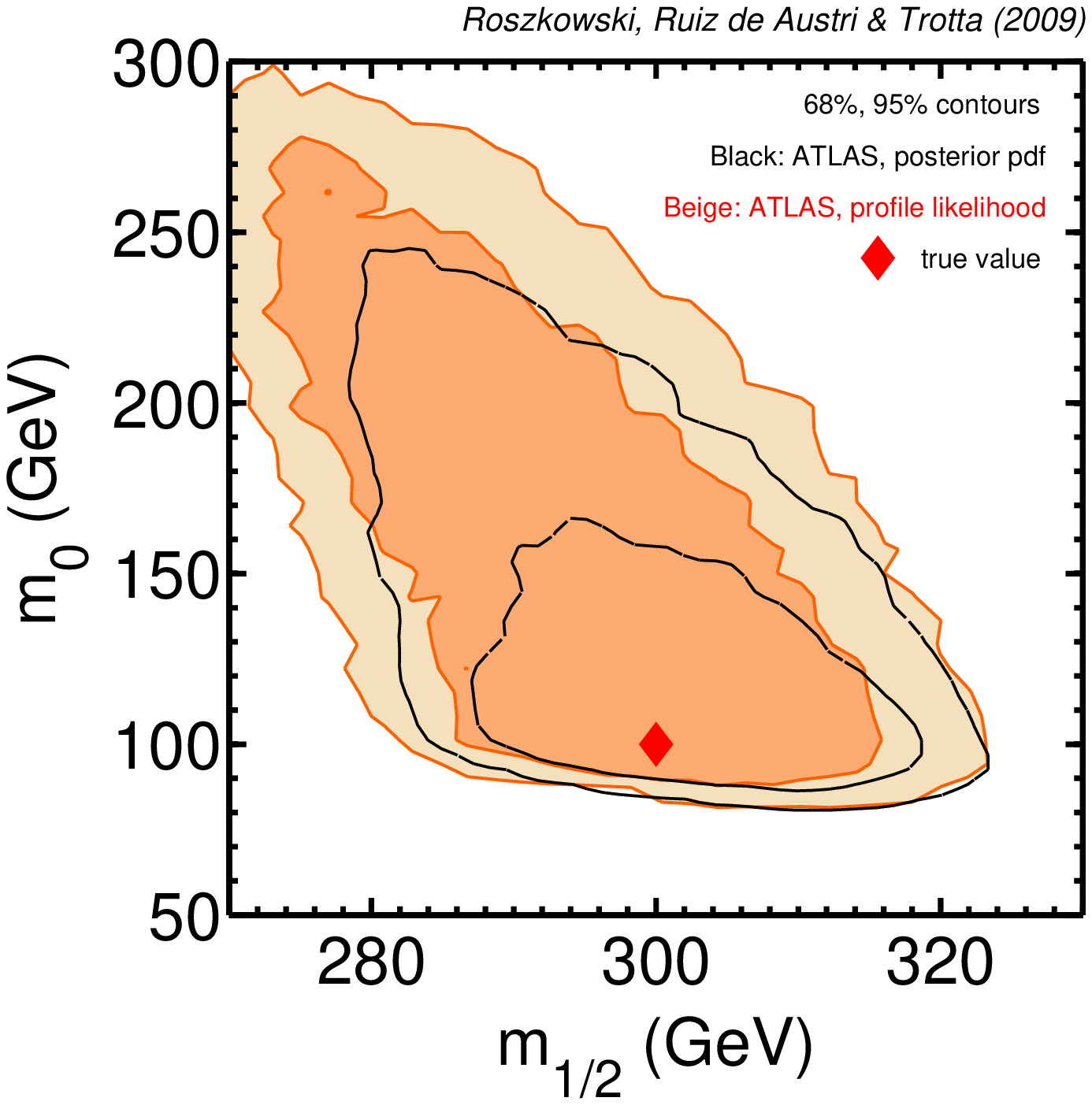}
\includegraphics[width=\ww]{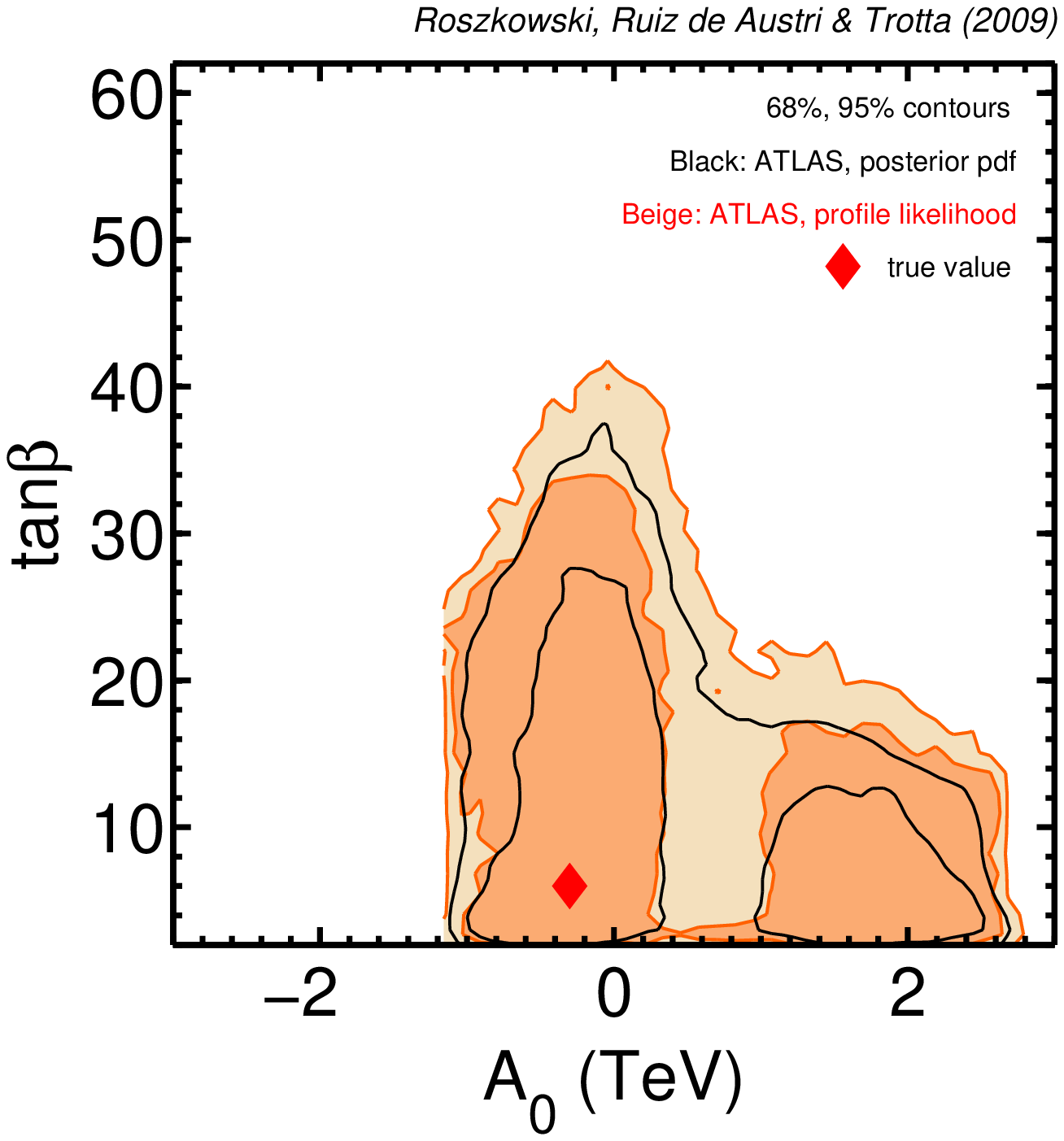}\\
\includegraphics[width=\ww]{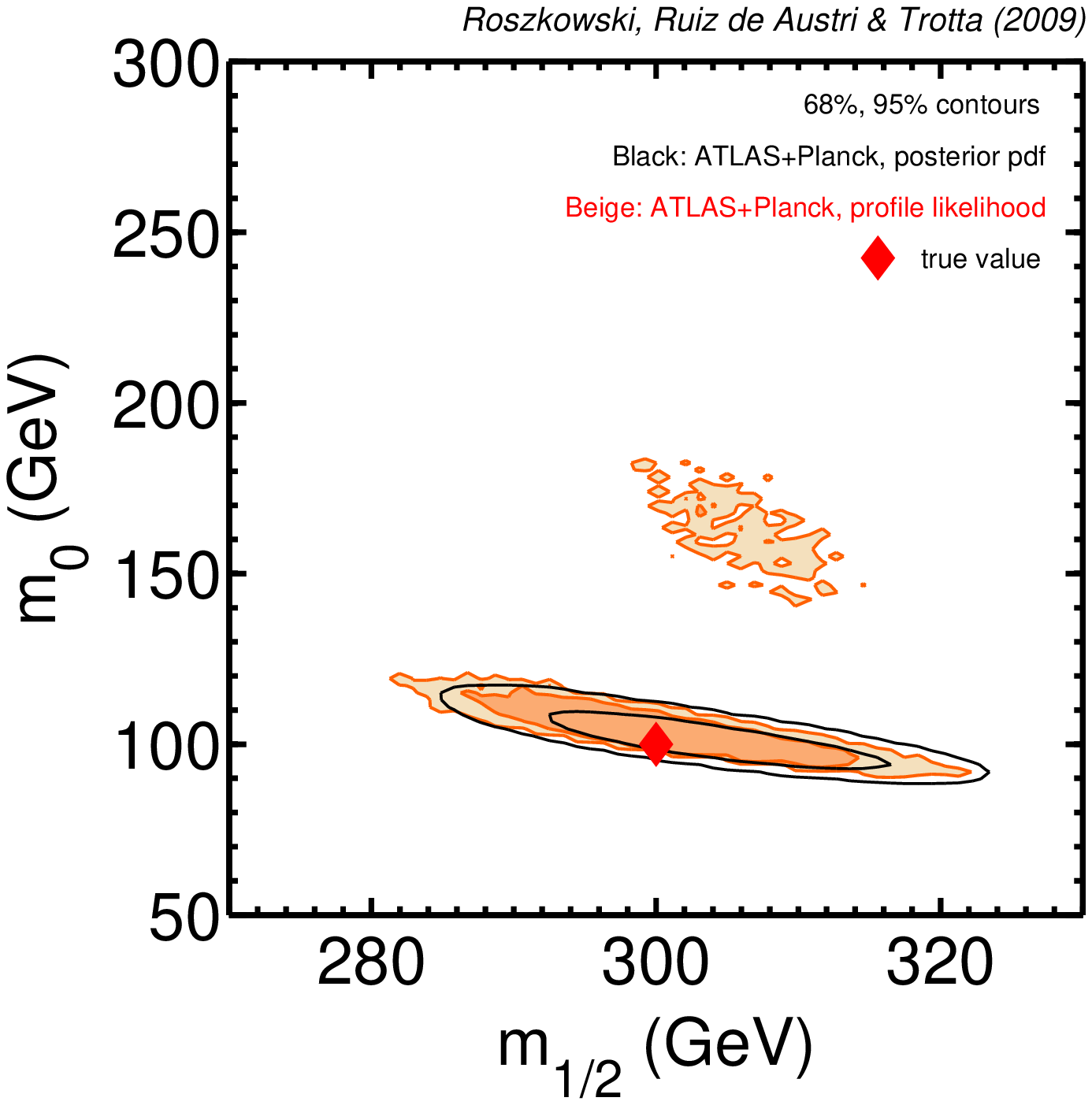}
\includegraphics[width=\ww]{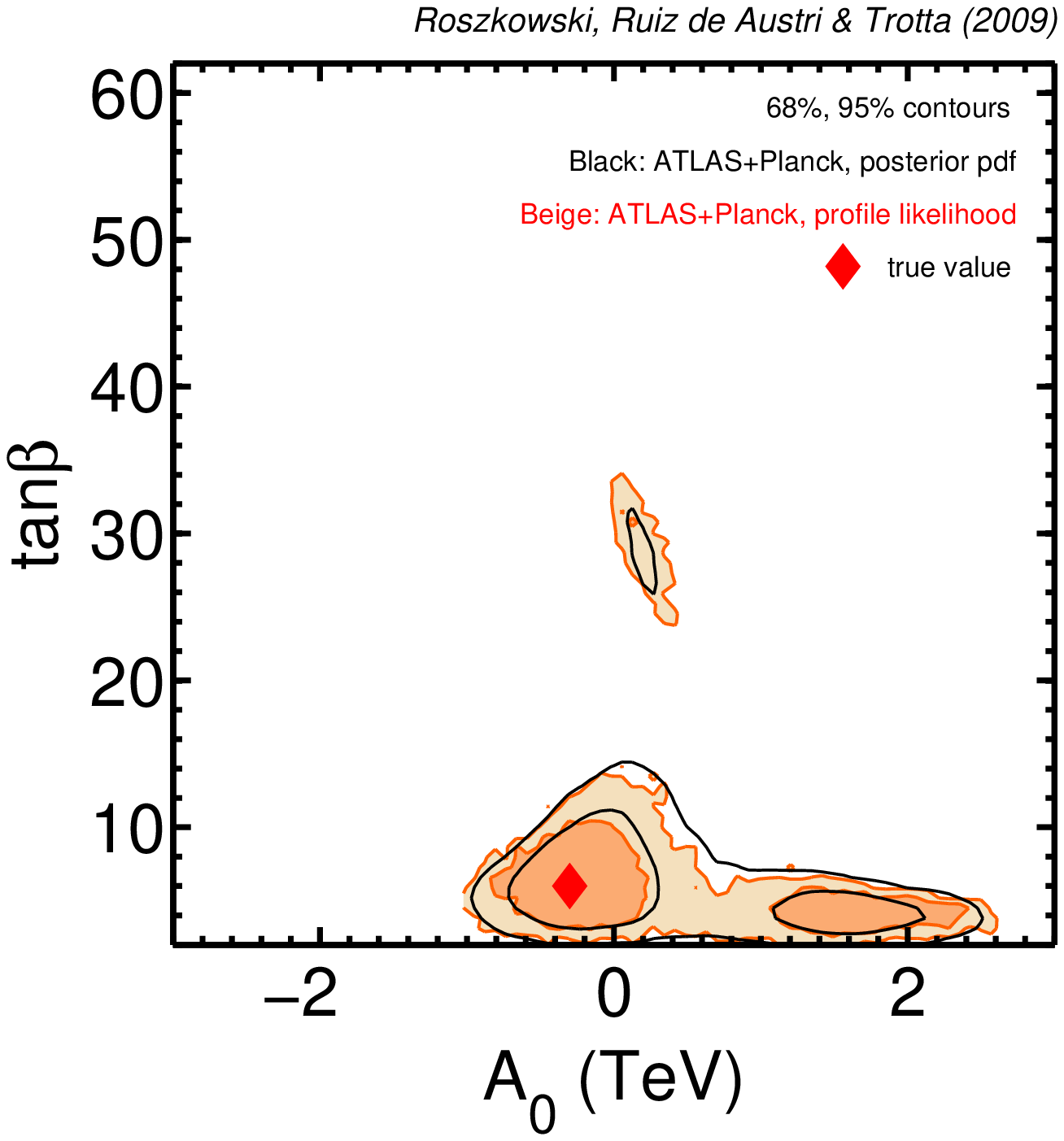} 
\caption[test]{Comparison between the profile likelihood (filled) and
  posterior (empty contours) using ATLAS mass spectrum data only (top
  row) and adding Planck (bottom row). With 
  this combination of data, the choice of statistics (Bayesian
  posterior or profile likelihood) becomes almost irrelevant, giving
  intervals that match at the 10\% level.} 
\label{fig:CMSSM_LHC_Planck_compare}
\end{center}
\end{figure}

In order to examine the robustness of
the results obtained with the Bayesian posterior pdf, in this Section
we compare them with what would be obtained by using a more
traditional $\chisq$-based analysis. We thus define the {\em profile
likelihood} for, \eg, the CMSSM parameter $\basis_1$, where $\basis_1,
\dots, \basis_8$ comprise the 4 CMSSM parameters and the 4 SM nuisance
parameters of Table~\ref{tab:su3input},
as 
\be \label{eq:profile}
\prof (\basis_1) \equiv \max_{\basis_2, \dots, \basis_8} \like(\data |
\basis), 
\ee 
where in our case $ \like(\data | \basis)$ is the full likelihood
function. Thus in the profile likelihood one maximises the value of
the likelihood along the other CMSSM and SM parameters, rather than
integrating it out as in the marginal posterior. From the profile
likelihood, confidence intervals are then obtained using the usual
likelihood-ratio criterion. In the context of MCMC scans of the
parameter space, the profile likelihood can be evaluated by simply
finding the maximum likelihood value within a given bin. This has been
studied before in the context of the CMSSM in
Refs.~\cite{bclw07,tfhrr1,GA}. Its interest lies in the fact that it
is a prior independent measure.  One should however be aware that,
given current data, the numerical value of the profile likelihood
remains dependent of the scanning algorithm employed, see the analysis
in Ref.~\cite{GA}. This problem is not relevant for the current paper,
as we are dealing with simulated data and we can thus double check
that our profile likelihood correctly peaks near the true maximum
likelihood value. Since we have found above little prior dependence of
the posterior pdf, this suggests that the posterior pdf is dominated
by the likelihood. Therefore we generically expect that the profile
likelihood will give similar statistical results as the posterior
studied above.

This is indeed confirmed in the top row of
\fig{fig:CMSSM_LHC_Planck_compare}, where we present the 68\% (inner
contours) and the 95\% confidence regions (outer contours) of the
profile likelihood for the ATLAS-only data case and we compare them
with the analogous regions derived from the posterior pdf presented
earlier in \fig{fig:CMSSM_LHC_2D_log}. We show only the log prior case
as the profile likelihood is prior independent (which we have verified
numerically). We can see that at the 95\%~\cl\ from the profile
likelihood is quite similar to the corresponding 95\% region derived
from the Bayesian posterior for all parameters, except for $\mzero$,
for which the profile likelihood yields looser constraint. It is worth
noticing that the posterior pdf yields a somewhat better
reconstruction of $\mzero$ and $\tanb$ and a similar one for the other
CMSSM parameters. However, the relative merits of the reconstructed
confidence regions from the posterior or from the profile likelihood
cannot be assessed here. It is in general a difficult task to decide
which statistics yields the ``best'' results (however one chooses to
define this). A possible way forward would be to carry out a coverage
study of the quoted confidence intervals, which is beyond the scope of
this paper.

On the other hand, what is encouraging is that, when the data becomes
sufficiently constraining, both statistical quantities produce
essentially equivalent confidence intervals. This is presented in
\fig{fig:CMSSM_LHC_Planck_compare} for the ATLAS+Planck case, which
should be compared with the bottom row of \fig{fig:DM_impact}.

\section{Summary and conclusions} \label{sec:summary}
In this paper we have examined prospects for reconstructing 
supersymmetric parameters from assumed future data that one can reasonably expect
to become available. To this end we focused on the Constrained MSSM
and on the benchmark point ATLAS SU3.

By following the ATLAS assumptions as closely as possible without
having access to the full simulated likelihood function, we arrived at
generally rather similar results for the reconstruction of the CMSSM
parameters, with the exception of $A_0$, for which our projected limit
appears somewhat weaker. We stress here that our method is generally
applicable, and that the quantitative discrepancies observed with the
ATLAS collaboration result are a consequence of the limited
information available about the precise shape of the likelihood
function. We therefore would urge experimentalists to make publicly
available numerical fits to the likelihood functions that could be
used to improved on the Gaussian assuption adopted here.  We
highlighted the computational advantage of our method which employs an
effective likelihood at the mass spectrum level, which allows to
shortcut the computationally expensive simulation of the whole
experimental setup. We also demonstrated that, once LHC data become
available, previously observed prior dependence of the results
disappears if one adopts the broad, non-informative flat
or log priors, although this may not be the case with any choice of
this class of priors.
We showed that the conclusions depend only mildly on which statistical
quantity one chooses to adopt, \ie, Bayesian posterior or profile
likelihood, in marked contrast with the present-day situation.  The
information from the likelihood can also be supplemented by a prior
encoding a preference for ``naturalness'', thus suppressing the
statistical weight of finely tuned regions. This choice leads to tighter
errors on $\mzero$ and $\tanb$, while hardly affecting the conclusions
on $\azero$ and $\mhalf$. We then extended the analysis by adding to
the likelihood function information about the neutralino dark matter
relic abundance by imposing WMAP-like and Planck-like
constraints. This improved the ability to reconstruct the value of
especially $\mzero$ and $\tanb$, much less so for $\mhalf$ (compared
to the ATLAS data only case), while the bi-modality in the
determination of $\azero$ could not be removed.

While the ATLAS SU3 point (and maybe also the CMSSM in the first
place) may be unlikely to be realized in Nature, the method presented here
appears to be powerful and robust enough to adequately reconstruct
supersymmetric parameters from summary statistics of LHC
measurements. The additional advantages presented here are the ability
to easily investigate several different theoretical scenarios with
relatively little computational effort, and the capability to produce
predictions for derived observable quantities, such as for example the
cosmological relic abundance and direct detection cross sections. The
inclusion of observational constraints from such probes has also been
demonstrated to be easily implemented. Finally, the favourable
scalability of our MultiNest scanning algorithm with the
dimensionality of the parameter space means that this method is in
principle ready to investigate theories with several tens of free
parameters, thereby opening the way to massive inference in
supersymmetry phenomenology.  As such we believe that our method will
be a useful tool to face the real data that is expected to soon start
arriving from the LHC, even if differs significantly from the case
considered here.

\medskip
{\bf Acknowledgements} \\ The authors would like to thank D.~Costanzo,
R.~Cousins, L.~Lyons and D.~Tovey for useful conversations, as well as C.~Topfel
and M.~Weber for providing the covariance matrix used in the ATLAS
Collaboration Report~\cite{atlas09}. A communication with K.~Desch,
M.~Uhlenbrock and P.~Wienemann is also acknowledged. We would like to thank an anonymous referee for useful comments. L.R. is
partially supported by STFC, the EC 6th Framework Programmes
MRTN-CT-2004-503369 and MRTN-CT-2006-035505.  The work of R.R. is
supported in part by MEC (Spain) under grant FPA2007-60323, by
Generalitat Valenciana under grant PROMETEO/2008/069 and by the
Spanish Consolider-Ingenio 2010 Programme CPAN (CSD2007-00042).
L.R. would like to thank the CERN Theory Division for hospitality
during the final stages of the project. R.T. would like to thank the
Galileo Galilei Institute for Theoretical Physics for the hospitality
and the INFN and the EU FP6 Marie Curie Research and Training Network
``UniverseNet'' (MRTN-CT-2006-035863) for partial support.


\begin{thebibliography}{99}

\bibitem{kkrw94}
G.~L.~Kane, C.~F.~Kolda, L.~Roszkowski and J.~D.~Wells,
\prd{49}{1994}{6173} [hep-ph/9312272].

\bibitem{sugra-refs}
A.~Chamseddine, R.~Arnowitt and P.~Nath, \prl{49}{1982}{970};
R.~Barbieri, S.~Ferrara and C.~Savoy, \plb{119}{1982}{343};
L.~J.~Hall, J.~Lykken and S.~Weinberg, \prd{27}{1983}{2359}; 
for a review, see, \eg, H.~P.~Nilles,
\prep{110}{1984}{1}. 

\bibitem{allanach-bayes}
B.~C.~Allanach and C.~G.~Lester,
\prd{73}{2006}{015013} [hep-ph/0507283]; 
B.~C.~Allanach, C.~G.~Lester and A.~M.~Weber,
\jhep{0612}{2006}{065} [hep-ph/0609295]; 
B.~C.~Allanach,
\plb{635}{2006}{123} [hep-ph/0601089].

\bibitem{rrtetal}
R.~Ruiz de Austri, R.~Trotta and L.~Roszkowski,
\jhep{0605}{2006}{002} [hep-ph/0602028]; 
L.~Roszkowski, R.~Ruiz de Austri and R.~Trotta, 
\jhep{0704}{2007}{084} [hep-ph/0611173] and
\jhep{0707}{2007}{075} [arXiv:0705.2012].


\bibitem{ehow06}
J.~R.~Ellis, \etal, 
\jhep{0605}{2006}{005} [hep-ph/0602220].

\bibitem{buchmueller08} O.~Buchmueller {\it et al.}, 
arXiv:0808.4128 [hep-ph].

\bibitem{ArkaniHamed:2005px}
  N.~Arkani-Hamed, \etal, 
\jhep{0608}{2006}{070} 
  [arXiv:hep-ph/0512190].

\bibitem{massreconstruction-refs}
  I.~Hinchliffe, \etal, 
\prd{55}{1997}{5520} 
  [hep-ph/9610544]; 
  C.~G.~Lester and D.~J.~Summers, 
\plb{463}{1999}{99} 
  [hep-ph/9906349]; 
%
%
%
%
%
%
%
%
%
  W.~S.~Cho,  \etal, 
\prl{100}{2008}{171801} 
  [arXiv:0709.0288]; 
 %
  G.~G.~Ross and M.~Serna, 
 \plb{665}{2008}{212} 
  [arXiv:0712.0943]; 
%
  M.~M.~Nojiri,  \etal, 
\jhep{0805}{2008}{014} 
  [arXiv:0712.2718];
 %
%
%
%
%
%
  H.~C.~Cheng,  \etal, 
\jhep{0712}{2007}{076} 
  [arXiv:0707.0030]. 
%
%

\bibitem{atlas09}
  G.~Aad, \etal,  [The ATLAS Collaboration],
  arXiv:0901.0512. 
  
\bibitem{softsusy}
B.~C.~Allanach,
\cpc{143}{2002}{305} [hep-ph/0104145].

\bibitem{micro} G. Belanger, \etal, 
\cpc{149}{2002}{103}
[hep-ph/0112278];
{\it MicrOMEGAs: Version 1.3, Comput. Phys. Commun. } {\bf 174}, 577
(2006) [hep-ph/0405253].

\bibitem{wmap5yr}
J.~Dunkley  \etal\ [The WMAP Collaboration],
\aps{180}{2009}{306} [arXiv:0803.0586].

\bibitem{Nojiri:2005ph}
  M.~M.~Nojiri,\etal, 
\jhep{0603}{2006}{063} 
[hep-ph/0512204].

\bibitem{planckdm} 
The Planck Collaboration, 
astro-ph/0604069.

\bibitem{Allanach:2004jh}
  B.~C.~Allanach, \etal, G.~Belanger, F.~Boudjema, A.~Pukhov and W.~Porod,
  hep-ph/0402161.

\bibitem{NS}
%
F.~Feroz and  M.~P. Hobson  
\mnras{384}{2008}{449}; 
F.~Feroz, \etal, 
arXiv:0809.3437. 



\bibitem{bclw07}
B.~C.~Allanach, \etal, 
\jhep{08}{2007}{023} [arXiv:0705.0487].


\bibitem{tfhrr1}
  R.~Trotta, \etal, 
\jhep{0812}{2008}{024}  [arXiv:0809.3792].

\bibitem{GA}
  Y.~Akrami, \etal, 
  JHEP {\bf 1004} (2010) 057
  [arXiv:0910.3950 [hep-ph]].
  
\bibitem{ccr08}
M.~E.~Cabrera, J.~A.~Casas and R.~R.~de Austri, 
  \jhep{0903}{2009}{075}
  [arXiv:0812.0536].
 
 \bibitem{Cabrera:2009dm}
  M.~E.~Cabrera, A.~Casas and R.~R.~de Austri,
  JHEP {\bf 1005} (2010) 043
  [arXiv:0911.4686 [hep-ph]].

\bibitem{Ellis:1986yg}
  J.~R.~Ellis, K.~Enqvist, D.~V.~Nanopoulos and F.~Zwirner,
  Mod.\ Phys.\ Lett.\  A {\bf 1} (1986) 57.
%
\bibitem{Barbieri:1987fn}
  R.~Barbieri and G.~F.~Giudice,
  Nucl.\ Phys.\  B {\bf 306} (1988) 63.
\bibitem{superbayes} 
See: \texttt{http://www.superbayes.org/} 
  
\bibitem{bbpw06} 
E.~A.~Baltz, \etal, 
\prd{74}{2006}{103521} [hep-ph/0602187].

\bibitem{Chan:1997bi}
  K.~L.~Chan, U.~Chattopadhyay and P.~Nath,
  {\it Naturalness, weak scale supersymmetry and the prospect for the  observation
  of supersymmetry at the Tevatron and at the LHC}, 
  \prd{58}{1998}{096004} [\hepph{9710473}].

\bibitem{focuspoint-fmm}
J.~L.~Feng, K.~T.~Matchev and T.~Moroi,
{\it Multi - TeV scalars are natural in minimal supergravity},
\prl{84}{2000}{2322} [\hepph{9908309}]
and
{\it  Focus points and naturalness in supersymmetry},
\prd{61}{2000}{075005} [\hepph{9909334}].


\end{thebibliography}
\end{document}